\documentclass[prd, twocolumn, superscriptaddress]{revtex4-2}
\usepackage{amsmath}
\usepackage{amssymb}

\usepackage{hyperref}

\usepackage{multirow}
\usepackage{array}

\usepackage{bm}

\usepackage[T1]{fontenc}

\allowdisplaybreaks

\begin{document}
\title{Localization in Quantum Field Theory for inertial and accelerated observers}

\author{Riccardo Falcone}
\affiliation{Department of Physics, University of Sapienza, Piazzale Aldo Moro 5, 00185 Rome, Italy}

\author{Claudio Conti}
\affiliation{Department of Physics, University of Sapienza, Piazzale Aldo Moro 5, 00185 Rome, Italy}

\begin{abstract}
We study the problem of localization in Quantum Field Theory (QFT) from the point of view of inertial and accelerated experimenters. We consider the Newton-Wigner, the Algebraic Quantum Field Theory (AQFT) and the modal localization schemes, which are, respectively, based on the orthogonality condition for states localized in disjoint regions of space, on the algebraic approach to QFT and on the representation of single particles as positive frequency solution of the field equation. We show that only the AQFT scheme obeys causality and physical invariance under diffeomorphisms.

Then, we consider the nonrelativistic limit of quantum fields in the Rindler frame. We demonstrate the convergence between the AQFT and the modal scheme and we show the emergence of the Born notion of localization of states and observables. Also, we study the scenario in which an experimenter prepares states over a background vacuum by means of nonrelativistic local operators and another experimenter carries out nonrelativistic local measurements in a different region. We find that the independence between preparation of states and measurements is not guaranteed when both experimenters are accelerated and the background state is different from Rindler vacuum, or when one of the two experimenters is inertial.
\end{abstract}

\maketitle

\section{Introduction}\label{Localization_in_accelerated_frame_Introduction}

The problem of localization in Quantum Field Theory (QFT) has been extensively discussed by different authors in the literature (for a brief review see, e.g., Ref.~\cite{localization_QFT}). Notably, Newton and Wigner \cite{RevModPhys.21.400} provided a unique notion of localization in relativistic quantum mechanics based on the orthogonality condition for states localized in disjoint spatial regions. However, as a consequence of the Hegerfeldt theorem \cite{PhysRevD.10.3320, PhysRevD.22.377}, the Newton-Wigner scheme appears to be in contrast with the notion of causality \cite{PhysRev.139.B963} and, hence, it is not suited for the description of relativistic local phenomena. The solution to this issue is given by the fundamental notion of localization that is based on the Algebraic Quantum Field Theory (AQFT) formalism \cite{10.1063/1.1704187, haag1992local, Halvorson:2006wj, Brunetti:2015vmh, 10.1007/978-3-030-38941-3_1}. At variance with the Newton-Wigner scheme, the AQFT scheme is characterized by nonlocal effects that originate from the quantum correlations in the Minkowski vacuum $| 0_\text{M} \rangle$ \cite{haag1992local, Redhead1995-REDMAA-2, PhysRevA.58.135} and are predicted by the Reeh-Schlieder theorem \cite{Reeh:1961ujh}. In particular, it has been noticed that nonunitary preparations of localized states influence measurements conducted in spacelike separated regions; by following Knight and Licht \cite{10.1063/1.1703731, 10.1063/1.1703925}, we say that nonunitarily locally prepared states are not always strictly localized. It has been argued that such a nonlocal effect does not lead to violations of causality \cite{Redhead1995-REDMAA-2, CLIFTON20011, VALENTE2014147, RevModPhys.90.045003}.

In all of the aforementioned works, the authors focused on the Minkowski spacetime and discussed the localization of states and observables from the point of view of inertial experimenters. In this manuscript, instead, we consider accelerated observes as well. The aim is to extend the problem of localization to include both inertial and accelerated frames. We use the formalism of Quantum Field Theory in Curved Spacetime (QFTCS) \cite{Wald:1995yp} to consider Rindler spacetimes in addition to Minkowski spacetimes and we adopt the algebraic approach to QFTCS \cite{haag1992local, Wald:1995yp} to relate localization schemes of one frame to the other.

\begin{table}
\begin{center}
\begin{tabular}{|l|| >{\centering\arraybackslash}m{7em} | >{\centering\arraybackslash}m{7em} | >{\centering\arraybackslash}m{7em} |}
\hline
& Experimenter A (preparation) & Experimenter B (observation) & Background vacuum state \\
\hline
\hline
ABM & inertial (Alice) & inertial (Bob) & Minkowski $|0_\text{M} \rangle$ \\
\hline
RaRbR & accelerated (Rachel) & accelerated (Rob) & Rindler $|0_\text{L},0_\text{R} \rangle$ \\
\hline
RaRbM & accelerated (Rachel) & accelerated (Rob) & Minkowski $|0_\text{M} \rangle$ \\
\hline
ARbM & inertial (Alice) & accelerated (Rob) & Minkowski $|0_\text{M} \rangle$ \\
\hline
\end{tabular}
\end{center}
\caption{Different measurement setups. The experimenter A prepares a local state over the background state and the experimenter B performs local measurement. Each experimenter may be inertial or accelerated and the background state can be either the Minkowski ($|0_\text{M} \rangle$) or the Rindler ($|0_\text{L},0_\text{R} \rangle$) vacuum.}\label{scenarios_Table}
\end{table}

We consider four different scenarios, which are summarized by Table \ref{scenarios_Table}. An experimenter (A) prepares states via local operations on the vacuum background $| \Omega \rangle$, whereas another experimenter (B) carries out measurements via local observables. Due to the nonequivalent notion of vacuum states in different frames \cite{Wald:1995yp}, here, we assume that $| \Omega \rangle$ may be equal to the Minkowski ($| 0_\text{M} \rangle$) or the Rindler ($| 0_\text{L}, 0_\text{R} \rangle$) vacuum \cite{RevModPhys.80.787}, depending on the situation. Each scenario is labeled by an acronym, identified by the nature of the experimenters A and B and of the background vacuum state $| \Omega \rangle$. The ABM scenario was already studied in literature since it is characterized by inertial experimenters (Alice and Bob) and by the Minkowski vacuum $|0_\text{M} \rangle$ as the background state. Conversely, the RaRbR and the RaRbM scenarios include only accelerated observers (Rachel and Rob); the difference between RaRbR and RaRbM is given by the background state $| \Omega \rangle$ which is equal to the Rindler ($|0_\text{L},0_\text{R} \rangle$) or the Minkowski ($|0_\text{M} \rangle$) vacuum, respectively. Finally, in the ARbM scenario, the experimenter A is inertial, whereas the experimenter B is accelerated; the local preparation of states is performed over the Minkowski vacuum $|0_\text{M} \rangle$.

To study the dependence between the outcomes of the measurements by the experimenter B and the local preparation of states by the experimenter A, we extend the definition of the Knight-Licht strict localization \cite{10.1063/1.1703731, 10.1063/1.1703925} to include the vacuum Rindler $|0_\text{L},0_\text{R} \rangle$ as a possible background state $| \Omega \rangle$ in addition to the Minkowski vacuum $|0_\text{M} \rangle$. We say that a state $| \psi \rangle $ is strictly localized in the spacetime region $\mathcal{O}_\text{A}$ over $| \Omega \rangle$ if it gives the same expectation values as $| \Omega \rangle$ for all measurements in any region $\mathcal{O}_\text{B}$ spacelike separated from $\mathcal{O}_\text{A}$, i.e.,
\begin{equation}\label{KnightLicht_property_Omega}
\langle \psi | \hat{O}_\text{B} | \psi \rangle = \langle \Omega | \hat{O}_\text{B}  | \Omega \rangle,
\end{equation}
with $\hat{O}_\text{B}$ as any observable localized in $\mathcal{O}_\text{B}$.

We consider different localization schemes to discuss the problem of localization for inertial and accelerated observers. We start from the Newton-Wigner scheme for massless scalar real fields in 1+1 Minkowski and Rindler spacetimes. By following the algebraic approach to QFTCS \cite{haag1992local, Wald:1995yp}, we show that the Minkowski-Newton-Wigner and the Rindler-Newton-Wigner schemes are incompatible to each other, in the sense that any operator that is localized with respect to one of the two schemes is not localized with respect to the other. It is known that the Newton-Wigner localization is not preserved by special relativistic transformations \cite{RevModPhys.21.400}; here, we find that it is not preserved by General Relativistic (GR) diffeomorphisms either. Also, we demonstrate that, in the Newton-Wigner ARbM scenario, the strict localization property is not guaranteed for unitarily prepared local states. In addition to the Hegerfeldt superluminal effect \cite{PhysRevD.10.3320, PhysRevD.22.377}, we find a violation of causality that is caused by the possibility for Alice to instantly send information to Rob via local unitary operations. These results give further motivations to disregard the Newton-Wigner scheme as a faithful description of local phenomena in the QFTCS regime.

After discussing the 1+1 Newton-Wigner scheme, we study the AQFT scheme in 3+1 dimensions for massive scalar real fields. In this case, the localization of states and observables is based on the definition of local field algebras $\mathfrak{A}(\mathcal{O})$ which are associated to spacetime regions $\mathcal{O}$. By using the fact that $\mathcal{O}$ is a frame independent object (i.e., without an intrinsic notion of coordinates), we prove that the Minkowski and the Rindler AQFT schemes are compatible, in the sense that inertial and accelerated observers always agree about the localization of states and observables. As a result, we show the covariant behavior of the AQFT localization under coordinate diffeomorphism. Also, due to the equivalence between the Minkowski and the Rindler AQFT schemes, we find that the RaRbR, RaRbM and ARbM scenarios share the same features of the ABM scenario. This also includes the lack of strict localization property in the case of selectively prepared local states \cite{10.1063/1.1703731, 10.1063/1.1703925, CLIFTON20011, VALENTE2014147, RevModPhys.90.045003}.

In addition to the AQFT scheme, we study the modal localization scheme, which is based on the representation of single particle states as positive frequency solutions of the respective Klein-Gordon equation \cite{Wald:1995yp, localization_QFT}. Due to the frame dependent content of particles \cite{Wald:1995yp, RevModPhys.80.787}, we find that the Minkowski and the Rindler modal schemes are incompatible to each other, in the sense that there is no state or observable that is simultaneously localized with respect to both schemes. Hence, the modal scheme does not satisfy the GR notion of physical equivalence between frames. Also, the strict localization property is not guaranteed for nonselective preparations of states. Such an issue occurs in both the Minkowski and the Rindler frame, leading to violations of relativistic causality in all four scenarios (i.e., ABM, RaRbR, RaRbM and ARbM). The noncovariant and acausal features make the modal scheme unsuitable for the description of local phenomena in the relativistic regime.

In the nonrelativistic limit \cite{PhysRevD.107.045012}, instead, the modal scheme acquires a genuine notion of localization due to the convergence to the AQFT scheme. In such a regime, we find the emergence of the Born interpretation of localized states, which is based on the factorization of the global Hilbert space into local Fock spaces and the global vacuum into the local vacua. Also, we show that the strict localization property is always satisfied in the nonrelativistic ABM and RaRbR scenarios. Conversely, in the nonrelativistic RaRbM and ARbM scenarios, it is only guaranteed for nonselective (e.g., unitary) preparations of states. In the RaRbM scenario, this is due to the fact that the background Minkowski vacuum $|0_\text{M} \rangle$ is entangled between the local Rindler-Fock spaces. In ARbM case, instead, the cause is the incompatibility between the Minkowski and the Rindler nonrelativistic schemes, which, in turn, is due to the frame dependent notion of the nonrelativistic limit \cite{PhysRevD.107.085016}.

The manuscript is organized as follows. In Sec.~\ref{NewtonWigner_localization_in_11_conformally_flat_spacetimes}, we study the Newton-Wigner scheme for Rindler massless scalar real fields in 1+1 dimensions. In Secs.~\ref{AQFT_localization_scheme_in_curved_spacetime} and \ref{Modal_localization_scheme_Rindler}, we consider Rindler scalar fields in 3+1 dimensions and we detail the AQFT and the modal scheme, respectively. A comparison between the two localization schemes is given in Sec.~\ref{Comparison_between_localization_schemes_Rindler}. Their nonrelativistic limit, instead, is detailed in Sec.~\ref{Localization_in_NRQFTCS}. Conclusions are drawn in Sec.~\ref{Localization_in_accelerated_frame_Conclusions}.

\section{Newton-Wigner scheme in 1+1 spacetime} \label{NewtonWigner_localization_in_11_conformally_flat_spacetimes}

The Newton-Wigner scheme in the 3+1 Minkowski spacetime is based on the definition of annihilation operators of particles with defined position as
\begin{equation}\label{a_NW}
\hat{a}_\text{NW}(\vec{x}) = \int_{\mathbb{R}^3} d^3 k \frac{e^{i \vec{k} \cdot \vec{x}}}{\sqrt{(2 \pi)^3 }} \hat{a}(\vec{k}),
\end{equation}
where $\hat{a}(\vec{k})$ is the annihilator of the particle with momentum $\vec{k}$ \cite{RevModPhys.21.400}. In the attempt to generalize definition of the Minkowski-Newton-Wigner scheme in curved spacetime, we immediately find a problem. Specifically, the existence of states with defined momentum are not always guaranteed, due to a possible lack of translational symmetry in the Lagrangian theory. 

Remarkably, massless scalar real fields in 1+1 spacetime are described by the same field equation in both the Minkowski and the Rindler frames. Indeed, the Klein-Gordon equation in 1+1 dimensions and with zero mass is
\begin{equation} \label{Klein_Gordon_11}
\left(  \partial_0^2 - c^2 \partial_1^2 \right) \hat{\phi}(t,x) = 0,
\end{equation}
whereas the Rindler-Klein-Gordon equation in 1+1 dimensions and with zero mass is
\begin{equation} \label{Rindler_Klein_Gordon_11}
\left(  \partial_0^2 - c^2 \partial_1^2 \right) \hat{\Phi}_\nu(T,X) = 0.
\end{equation}
Here, $\hat{\phi}(t,x)$ is the scalar field in the Minkowski frame, which is represented by the coordinates $(t,x)$, whereas $\hat{\Phi}_\nu(T,X)$ is the scalar field in the $\nu$-Rindler frame with coordinates $(T,X)$. The variable $\nu$ represents the Minkowski wedge covered by the respective Rindler coordinate system $(T,X)$; the left ($\nu = \text{L}$) and right ($\nu = \text{R}$) wedges are defined, respectively, by $x < c |t|$ and $x > c |t|$. The coordinate transformation relating one frame to the other is given by $t = t_\nu (T, X)$ and $x = x_\nu (T, X)$, with
\begin{subequations}\label{Rindler_coordinate_transformation_11}
\begin{align}
& t_\nu(T,X) = \frac{e^{s_\nu a X}}{c a} \sinh(c a T), \\
 & x_\nu(T,X) = s_\nu \frac{e^{s_\nu a X}}{a} \cosh(c a T), 
\end{align}
\end{subequations}
and where $s_\nu$ is such that $s_\text{L} = -1$ and $s_\text{R} = 1$. The parameter $a$ is proportional to the acceleration of the Rindler frame.

As a consequence of Eqs.~(\ref{Klein_Gordon_11}) and (\ref{Rindler_Klein_Gordon_11}), both the Minkowski and the Rindler fields appear to be symmetric with respect to spatial translations and admit the free modes
\begin{equation}\label{f_11}
f(k,t,x) = \sqrt{\frac{\hbar}{4\pi c |k|}}  e^{-i c |k|t+ikx}
\end{equation}
as solutions of the respective Klein-Gordon equation. This leads to the decomposition
\begin{subequations} 
\begin{align}
\hat{\phi}(t,x) = & \int_{\mathbb{R}} dk  \left[ f(k,t,x) \hat{a}(k) + f^*(k,t,x) \hat{a}^\dagger (k) \right],\label{scalar_field_M}\\
 \hat{\Phi}_\nu(T,X) = & \int_{\mathbb{R}} dK \left[ f(K,T,X) \hat{A}_\nu(K) \right. \nonumber \\
& \left. + f^*(K,T,X) \hat{A}_\nu^\dagger(K) \right],\label{scalar_field_R}
\end{align}
\end{subequations}
with $\hat{a}(k)$ and $\hat{A}_\nu(K)$ as, respectively, the annihilator of the Minkowski particle with momentum $k$ and the annihilator of the $\nu$-Rindler particle with momentum $K$. The respective canonical commutation relations are
\begin{align}\label{a_commutation}
& [\hat{a}(k), a^\dagger(k') ] = \delta(k-k'), & [\hat{a}(k), \hat{a}(k') ] = 0
\end{align}
and
\begin{subequations}\label{A_commutation} 
\begin{align}
& [\hat{A}_\nu(K), \hat{A}_{\nu'}^\dagger(K') ] = \delta_{\nu\nu'} \delta(K-K'),\\
 & [\hat{A}_{\nu}(K), \hat{A}_{\nu'}(K') ] = 0.
\end{align}
\end{subequations} 
In this case, the Minkowski vacuum $|0_\text{M} \rangle$ is defined by $\hat{a}(k) |0_\text{M} \rangle = 0$ for any $k \in \mathbb{R}$, whereas the Rindler vacuum $|0_\text{L},0_\text{R} \rangle$ by $\hat{A}_\nu(K) |0_\text{M} \rangle = 0$ for any $\nu \in \{ \text{L}, \text{R} \}$ and for any $K \in \mathbb{R}$.

By following Newton and Wigner \cite{RevModPhys.21.400}, we define the Minkowski-Newton-Wigner annihilation operator $\hat{a}_\text{NW}(x)$ as the anti-Fourier transform of $\hat{a}(k)$, i.e.,
\begin{equation}\label{a_NW_11}
\hat{a}_\text{NW}(x) = \int_{\mathbb{R}} d k \frac{e^{i k x}}{\sqrt{2 \pi}} \hat{a}(k).
\end{equation}
Analogously, we define the Rindler-Newton-Wigner operator $\hat{A}_{\text{NW},\nu}(X)$ as
\begin{equation}\label{A_NW_X_A_K}
\hat{A}_{\text{NW},\nu}(X) = \int_{\mathbb{R}} dK \frac{e^{i K X}}{\sqrt{2\pi}} \hat{A}_\nu (K).
\end{equation}

The operator $\hat{a}_\text{NW}(x)$ with fixed $x$ generated the local algebra $\mathfrak{A}_\text{M}^\text{NW}(x)$ in the Minkowski coordinate point $x$ with respect to the Minkowski-Newton-Wigner scheme. More generally, for any $\mathcal{V} \subset \mathbb{R}$, the operators $\hat{a}_\text{NW}(x)$ with $x \in \mathcal{V}$ generate the local algebra $\mathfrak{A}_\text{M}^\text{NW}(\mathcal{V})$ in the coordinate region $\mathcal{V}$. Any operator $\hat{O}$ is localized in $\mathcal{V}$ with respect to the Minkowski-Newton-Wigner scheme if it is an element of $\mathfrak{A}_\text{M}^\text{NW}(\mathcal{V})$. The notion of local operators gives the definition of localized states $| \psi \rangle$ based on the local preparation of $| \psi \rangle$ over the background $| \Omega \rangle$. We say that the state $| \psi \rangle$ is localized in $\mathcal{V}$ over the background $|\Omega \rangle $  with respect to the Minkowski-Newton-Wigner scheme if there is a local operator $\hat{O} \in \mathfrak{A}_\nu^{\text{NW}}(\mathcal{V})$ such that $| \psi \rangle = \hat{O} |\Omega \rangle $.

By using the Rindler operators $\hat{A}_\nu(K)$ instead of the Minkowski operators $\hat{a}(k)$, we may extend the definition of the Newton-Wigner scheme to the Rindler spacetime. In that case, the operators $\hat{A}_{\text{NW},\nu}(X)$ generate the local algebras $\mathfrak{A}_\nu^{\text{NW}}(X)$ in wedges $\nu = \{ \text{L}, \text{R} \}$ and space points $X \in \mathbb{R}$. Similarly, we define the local algebras $\mathfrak{A}_\nu^{\text{NW}}(\mathcal{V})$ in the extended space regions $\mathcal{V} \subset \mathbb{R}$. The operator $ \hat{O}$ is said to be Rindler-Newton-Wigner localized in the wedge $\nu$ and the region $\mathcal{V}$ if $\hat{O} \in \mathfrak{A}_\nu^{\text{NW}}(\mathcal{V})$. The state $| \psi \rangle = \hat{O} |\Omega \rangle $ is Rindler-Newton-Wigner localized in $\nu$ and $\mathcal{V}$ over the background $|\Omega \rangle $ if $\hat{O} \in \mathfrak{A}_\nu^{\text{NW}}(\mathcal{V})$.

In the following subsections we study the ABM, the RaRbR, the RaRbM and the ARbM scenarios introduced in Sec.~\ref{Localization_in_accelerated_frame_Introduction}.

\subsection{Newton-Wigner ABM scenario}\label{NewtonWigner_ABM_scenario}

The Newton-Wigner ABM scenario is already discussed in the literature (see, e.g., Ref.~\cite{localization_QFT}), as it is exclusively described by means of the Newton-Wigner scheme in Minkowski spacetime. States and observables are, respectively, prepared and measured by inertial experimenters and the background state is the Minkowski vacuum $| 0_\text{M} \rangle$. In this subsections, we detail the most relevant features of the theory.

By using Eqs.~(\ref{a_commutation}) and (\ref{a_NW_11}), one obtains the commutation relation
\begin{subequations}\label{a_NW_commutation}
\begin{align}
& [\hat{a}_\text{NW}(x), \hat{a}_\text{NW}^\dagger(x') ] =  \delta(x-x'), \\
& [\hat{a}_\text{NW}(x), \hat{a}_\text{NW}(x') ] = 0.
\end{align}
\end{subequations}
Equation (\ref{a_NW_commutation}) implies that the Minkowski-Newton-Wigner operators $\hat{a}_\text{NW}(x)$ are local annihilation operators, which induce a factorization of the global Minkowski-Fock space $\mathcal{H}_\text{M}$ into local Fock spaces $ \mathcal{H}_\text{M}^\text{NW}(\mathcal{V}_i)$, where $\{ \mathcal{V}_i \}$ is any partition of $\mathbb{R}$. By denoting the vacuum state of each Fock space $\mathcal{H}_\text{M}^\text{NW}(\mathcal{V}_i)$ as $| 0_{\text{M}}^\text{NW}(\mathcal{V}_i) \rangle$, we find that the Minkowski vacuum $| 0_\text{M} \rangle$ factorizes into the local vacua, i.e., $| 0_{\text{M}} \rangle = \bigotimes_i | 0_{\text{M}}(\mathcal{V}_i) \rangle$. 

Due to the factorizations $\mathcal{H}_\text{M} = \bigotimes_i \mathcal{H}_\text{M}^\text{NW}(\mathcal{V}_i)$ and $| 0_{\text{M}} \rangle = \bigotimes_i | 0_{\text{M}}(\mathcal{V}_i) \rangle$, any Minkowski-Fock state that is localized in $\mathcal{V}_\text{A}$ over the Minkowski vacuum $| 0_{\text{M}} \rangle$ factorizes as
\begin{equation}\label{NW_localization_decomposition}
| \psi \rangle = \hat{O}_\text{A} | 0_{\text{M}}(\mathcal{V}) \rangle \otimes \left[ \bigotimes_i | 0_{\text{M}}(\mathcal{V}_i) \rangle \right],
\end{equation}
where, in this case, $\{ \mathcal{V}_i \}$ is a partition of $\mathbb{R} \setminus \mathcal{V}_\text{A}$ and $\hat{O}_\text{A}$ is an operator acting on $ \mathcal{H}_\text{M}^\text{NW}(\mathcal{V}_\text{A})$. Equation (\ref{NW_localization_decomposition}) implies that the strict localization property is always satisfied in the Minkowski-Newton-Wigner scheme. Explicitly, this means that any state $| \psi \rangle = \hat{O}_\text{A}  | 0_\text{M} \rangle $ localized in $\mathcal{V}_\text{A}$ with respect to the Minkowski-Newton-Wigner scheme over the Minkowski vacuum $| 0_{\text{M}} \rangle$ satisfies Eq.~(\ref{KnightLicht_property_Omega}) with $| \Omega \rangle = | 0_{\text{M}} \rangle$ and for any observable $\hat{O}_\text{B} \in \mathfrak{A}_\text{M}^\text{NW}(\mathcal{V}_\text{B})$ with $\mathcal{V}_\text{B}$ disjoint from $\mathcal{V}_\text{A}$. In other words, $| \psi \rangle$ always appears indistinguishable from $| 0_\text{M} \rangle $ in any $\mathcal{V}_\text{B}$ disjoint from $\mathcal{V}_\text{A}$. As a result, we find that, in the ABM scenario, the outcomes of Bob's measurements are always independent of the preparation of the state by Alice.

\subsection{Newton-Wigner RaRbR scenario}

At variance with Sec.~\ref{NewtonWigner_ABM_scenario}, here we consider the Rindler-Newton-Wigner scheme to define localized states and observables from the point of view of accelerated experimenters. Also, we choose the Rindler vacuum $|0_\text{L},0_\text{R} \rangle$ as the background over which states are prepared, i.e., $|\Omega \rangle = |0_\text{L},0_\text{R} \rangle$. 

By definition, the Rindler-Newton-Wigner operator $\hat{A}_{\text{NW},\nu}(X)$ annihilates the vacuum $|0_\text{L},0_\text{R} \rangle$ and has the same algebraic properties as $\hat{a}_\text{NW}(x)$ from the Minkowski spacetime. Hence, all the features of the Minkowski-Newton-Wigner scheme detailed in Sec.~\ref{NewtonWigner_ABM_scenario} also apply here.

In particular, we find that the Rindler-Newton-Wigner operators satisfy the canonical commutation relation
\begin{subequations}\label{A_NW_commutation}
\begin{align}
& [\hat{A}_{\text{NW},\nu}(X), \hat{A}_{\text{NW},\nu'}^\dagger(X') ] = \delta_{\nu\nu'} \delta(X-X'),\\
 & [\hat{A}_{\text{NW},\nu}(X), \hat{A}_{\text{NW},\nu'}(X') ] = 0.
\end{align}
\end{subequations}
As a consequence of Eq.~(\ref{A_NW_commutation}), the global Rindler-Fock space $\mathcal{H}_{\text{L},\text{R}}$ factorizes into local Rindler-Fock spaces $\mathcal{H}_\nu^\text{NW}(\mathcal{V}_i)$ upon which elements of $ \mathfrak{A}_\nu^{\text{NW}}(\mathcal{V}_i)$ act. We note by $| 0_\nu (\mathcal{V}) \rangle$ the vacuum of $\mathcal{H}_\nu^\text{NW}(\mathcal{V})$. The factorization of the global Rindler vacuum $|0_\text{L},0_\text{R} \rangle$ in $\mathcal{H}_{\text{L},\text{R}} = \bigotimes_\nu \bigotimes_i \mathcal{H}_\nu^\text{NW}(\mathcal{V}_i)$ is $|0_\text{L},0_\text{R} \rangle = \bigotimes_\nu \bigotimes_i | 0_\nu (\mathcal{V}_i) \rangle$.

Due to the factorization of the Rindler-Fock space $\mathcal{H}_{\text{L},\text{R}} = \bigotimes_\nu \bigotimes_i \mathcal{H}_\nu^\text{NW}(\mathcal{V}_i)$ and the Rindler vacuum $|0_\text{L},0_\text{R} \rangle = \bigotimes_\nu \bigotimes_i | 0_\nu (\mathcal{V}_i) \rangle$, we find that for any couple of operators $\hat{O}_\text{A} \in \mathfrak{A}_{\nu_\text{A}}^{\text{NW}}(\mathcal{V}_\text{A})$ and $\hat{O}_\text{B} \in \mathfrak{A}_{\nu_\text{B}}^{\text{NW}}(\mathcal{V}_\text{B})$, the local state $| \psi \rangle = \hat{O}_\text{A} |0_\text{L},0_\text{R} \rangle $ satisfies Eq.~(\ref{KnightLicht_property_Omega}) with $| \Omega \rangle = |0_\text{L},0_\text{R} \rangle$ if $\nu_\text{A} \neq \nu_\text{B}$ or if $\mathcal{V}_\text{A}$ and $\mathcal{V}_\text{B}$ are disjoint. Hence, any state localized with respect to the Rindler-Newton-Wigner scheme is also strictly localized over the Rindler vacuum.

This result can be understood in terms of the Newton-Wigner RaRbR scenario introduced in Sec.~\ref{Localization_in_accelerated_frame_Introduction}. An accelerated experimenter (Rachel) prepares the state $| \psi \rangle$ over the Rindler vacuum $|0_\text{L},0_\text{R} \rangle $ by means of Newton-Wigner local operators in $\mathfrak{A}_{\nu_\text{A}}^{\text{NW}}(\mathcal{V}_\text{A})$. The accelerated observer Rob performs measurements by using elements of $\mathfrak{A}_{\nu_\text{B}}^{\text{NW}}(\mathcal{V}_\text{B})$. We find that the local preparation of $| \psi \rangle$ in $\nu_\text{A}$ and $\mathcal{V}_\text{A}$ do not influence measurements in $\nu_\text{B}$ and $\mathcal{V}_\text{B}$. This result is equivalent to the one obtained for the Newton-Wigner ABM scenario in Sec.~\ref{NewtonWigner_ABM_scenario}.

\subsection{Newton-Wigner RaRbM scenario}\label{RaRbM_scenario}

In this subsection, we still assume that all experimenters are accelerated. However, we consider the case in which laboratory phenomena are described as occurring over the Minkowski vacuum background $| 0_\text{M} \rangle$.

The identity relating $| 0_\text{M} \rangle$ to the Rindler vacuum $|0_\text{L},0_\text{R} \rangle$ can be obtained by computing the Bogoliubov transformation between the Minkowski and the Rindler operators \cite{RevModPhys.80.787}
\begin{align}\label{Bogolyubov_transformation}
\hat{a}(k) = & \int_{\mathbb{R}}dK \left[ \alpha(k,K)\hat{A}_\text{L}(K) - \beta^*(k,K) \hat{A}^\dagger_\text{L}(K) \right. \nonumber \\
& \left. + \alpha^*(k,K) \hat{A}_\text{R}(K) - \beta(k,K) \hat{A}^\dagger_\text{R}(K) \right],
\end{align}
with Bogoliubov coefficients
\begin{subequations}\label{Bogolyubov_coefficients}
\begin{align}
& \alpha(k,K) = \theta(kK) \sqrt{\frac{K}{k}}F(k,K), \\
 & \beta(k,K) = \theta(kK) \sqrt{\frac{K}{k}}F(-k,K)
\end{align}
\end{subequations}
and with
\begin{equation}\label{F_k_K}
F(k,K) = \int_{\mathbb{R}}  \frac{dX}{2 \pi}  \exp \left( - i K X + i \frac{k}{a} e^{a X} \right),
\end{equation}
Equation (\ref{Bogolyubov_transformation}) leads to
\begin{equation} \label{SS_S}
| 0_\text{M} \rangle \propto \exp \left( \int_\mathbb{R} dK e^{-\pi |K| / a} \hat{A}^\dagger_\text{L}(K) \hat{A}^\dagger_\text{R}(K) \right) | 0_\text{L}, 0_\text{R} \rangle.
\end{equation}

As a consequence of Eq.~(\ref{SS_S}), we find that the factorization $|0_\text{L},0_\text{R} \rangle = \bigotimes_\nu \bigotimes_i | 0_\nu (\mathcal{V}_i) \rangle$ does not hold for the Minkowski vacuum, in the sense that $|0_\text{M} \rangle \neq \bigotimes_\nu \bigotimes_i | 0_\nu (\mathcal{V}_i) \rangle$. This implies that Eq.~(\ref{KnightLicht_property_Omega}) does not necessarily hold when $| \Omega \rangle = |0_\text{M} \rangle$, even if $| \psi \rangle = \hat{O}_\text{A} | 0_\text{M} \rangle$ and if the operators $\hat{O}_\text{A} $ and $\hat{O}_\text{B}$ are Rindler-Newton-Wigner localized in two disjoint regions. 

However, one can show that the unitarity preparation of $| \psi \rangle $ is a sufficient condition for Eq.~(\ref{KnightLicht_property_Omega}). In particular, if $\hat{O}_\text{A}$ is a unitary element of $\mathfrak{A}_{\nu_\text{A}}^{\text{NW}}(\mathcal{V}_\text{A})$ and if $\hat{O}_\text{B}$ is an element of $\mathfrak{A}_{\nu_\text{B}}^{\text{NW}}(\mathcal{V}_\text{B})$, with $\nu_\text{B} \neq \nu_\text{A}$ or with $\mathcal{V}_\text{B}$ disjoint from $\mathcal{V}_\text{A}$, then we have that
\begin{equation}\label{local_unitary_operator_measurament_A}
\langle 0_\text{M} | \hat{O}_\text{A}^\dagger \hat{O}_\text{B} \hat{O}_\text{A} | 0_\text{M} \rangle = \langle 0_\text{M} | \hat{O}_\text{A}^\dagger \hat{O}_\text{A} \hat{O}_\text{B} | 0_\text{M} \rangle = \langle 0_\text{M} | \hat{O}_\text{B} | 0_\text{M} \rangle.
\end{equation}
since $[\hat{O}_\text{A}, \hat{O}_\text{B}] = 0$ and $\hat{O}_\text{A}^\dagger \hat{O}_\text{A} = 1$. As a result, we find that when the Minkowski vacuum $|0_\text{M} \rangle$ is chosen as a background state, the strict localization property in the Rindler-Newton-Wigner scheme is only guaranteed for unitary preparations of states.

Notice that the unitary condition $\hat{O}_\text{A}^\dagger \hat{O}_\text{A} = 1$ has been crucial for the proof of Eq.~(\ref{KnightLicht_property_Omega}). This result can be extended to a larger class of operations, known as local nonselective operations, which are defined by means of local Kraus operators $\hat{K}_i \in \mathfrak{A}_{\nu_\text{A}}^{\text{NW}}(\mathcal{V}_\text{A})$ \cite{kraus1983states} satisfying the nonselective condition $\sum_i \hat{K}_i^\dagger \hat{K}_i = 1$. The state prepared by $\hat{K}_i$ is described by the statistical operator
\begin{equation}
\hat{\rho} = \frac{\sum_i \hat{K}_i |0_\text{M} \rangle \langle 0_\text{M} | \hat{K}_i^\dagger}{\sum_i \langle 0_\text{M} | \hat{K}_i^\dagger \hat{K}_i |0_\text{M} \rangle}.
\end{equation}
Notice that unitary transformations are nonselective operations with a single Kraus operator. For any $\hat{O}_\text{B} \in \mathfrak{A}(\mathcal{O}_\text{B})$ spacelike separated from $\mathcal{O}_\text{A}$, we find that
\begin{equation}\label{Tr_rho_A_nonselective_0M}
\text{Tr} (\hat{\rho}\hat{O}_\text{B}) = \langle 0_\text{M}| \hat{O}_\text{B} |0_\text{M} \rangle
\end{equation}
as a consequence of the commutation identity  $[\hat{K}_i, \hat{O}_\text{B}] = 0$ and the nonselective condition $\sum_i \hat{K}_i^\dagger \hat{K}_i = 1$. In Eq.~(\ref{Tr_rho_A_nonselective_0M}), the outcome of the measurements carried out by Rob are given by the Minkowski vacuum $|0_\text{M} \rangle$, which means that $\hat{\rho}$ is strictly localized in $\mathcal{O}_\text{A}$.

We can interpret this results in the context of the RaRbM scenario, which is equivalent to the RaRbR scenario, except for the Minkowski vacuum $| 0_\text{M} \rangle$ replacing the Rindler vacuum $|0_\text{L},0_\text{R} \rangle$ as the background state. We find that, in the Newton-Wigner RaRbM experiment, selective preparations by Rachel may be detected by Rob; whereas nonselective operations on the vacuum $|0_\text{M} \rangle$ suffice to not influence measurements in the other disjoint region.

The effect looks very similar to the Reeh-Schlieder nonlocality \cite{10.1063/1.1703731, 10.1063/1.1703925, CLIFTON20011, VALENTE2014147, RevModPhys.90.045003} for two reasons: (i) The local preparation of states affects local measurements in a separated region, except when only nonselective preparations are considered; (ii) The origin of the effect is ascribed to a background state that does not factorize into local vacua.

\subsection{Newton-Wigner ARbM scenario}\label{NewtonWigner_ARbM_scenario}

The last scenario that we want to detail is the ARbM. In this case, states are prepared by an inertial observer (Alice) via Minkowski-Newton-Wigner operators $\hat{a}_\text{NW}(x)$.

The identity relating $\hat{a}_\text{NW}(x)$ to $\hat{A}_{\text{NW},\nu}(X)$ can be computed by means of Eqs.~(\ref{a_NW_11}), (\ref{A_NW_X_A_K}) and (\ref{Bogolyubov_transformation}), which give
\begin{align}\label{NW_operators_MR}
& \hat{a}_\text{NW}(x)\nonumber \\
 = & \int_{\mathbb{R}} dX  \left[  \tilde{\alpha}_\text{L}(x,X)  \hat{A}_\text{NW,L}(X) - \tilde{\beta}_\text{L}^*(x,X) \hat{A}^\dagger_\text{NW,L}(X) \right. \nonumber \\
& \left. +  \tilde{\alpha}_\text{R}^*(x,X) \hat{A}_\text{NW,R}(X) - \tilde{\beta}_\text{R}(x,X) \hat{A}^\dagger_\text{NW,R}(X) \right],
\end{align} 
with
\begin{subequations}\label{alpha_tilde_beta_tilde}
\begin{align}
& \tilde{\alpha}_\nu(x,X) = \int_{\mathbb{R}} dk \int_{\mathbb{R}} dK \frac{e^{-s_\nu i k x + s_\nu i K X}}{2\pi} \alpha(k,K), \\
& \tilde{\beta}_\nu(x,X) = \int_{\mathbb{R}} dk \int_{\mathbb{R}} dK \frac{e^{s_\nu i k x + s_\nu i K X}}{2\pi} \beta(k,K).
\end{align}
\end{subequations}

By using Eqs.~(\ref{Bogolyubov_coefficients}) and (\ref{F_k_K}) in Eq.~(\ref{alpha_tilde_beta_tilde}), we find that the support of the kernel functions $\tilde{\alpha}_\nu(x,X)$ and $\tilde{\beta}_\nu(x,X)$ with respect to the variable $X$ is the entire real axis $\mathbb{R}$. Hence, the Minkowski-Newton-Wigner operator $\hat{a}_\text{NW}(x)$ is not localized with respect to the Rindler-Newton-Wigner scheme. Due to the nontrivial Bogoliubov transformation relating the two sets of creators and annihilators, we find that the Minkowski and the Rindler Newton-Wigner schemes are incompatibles. Hence, the notion of Newton-Wigner localization is noncovariant with respect to GR diffeomorphisms. 

In Sec.~\ref{NewtonWigner_ABM_scenario}, we discussed the Newton-Wigner ABM scenario, where both preparation of states and measurements are preformed by means of operators localized with respect to the Minkowski-Newton-Wigner scheme and over the Minkowski vacuum $| 0_\text{M} \rangle$. The independence between preparation and measurement is a consequence of the factorization of the global Minkowski-Fock space $\mathcal{H}_\text{M}$ into local Fock spaces $\mathcal{H}_\text{M}^\text{NW}(\mathcal{V}_i)$ with local vacua $| 0_\text{M} (\mathcal{V}_i)\rangle$ and the global vacuum $| 0_\text{M} \rangle$ into the local vacua $| 0_\text{M} (\mathcal{V}_i)\rangle$. In the ARbM scenario, the inertial observer Bob is replaced by the accelerated observer Rob, who has access to the algebra $\mathfrak{A}_{\nu_\text{B}}^\text{NW} (\mathcal{V}_\text{B})$. Due to the incompatibility between the Minkowski and the Rindler Newton-Wigner schemes, any operator $\hat{O}_\text{B} \in \mathfrak{A}_{\nu_\text{B}}^\text{NW} (\mathcal{V}_\text{B})$ measured by Rob is actually global with respect to the Minkowski-Newton-Wigner scheme, i.e., $\hat{O}_\text{B} \in \mathfrak{A}_\text{M}^\text{NW} (\mathbb{R})$. For this reason, the preparation of the state $| \psi \rangle = \hat{O}_\text{A} |0_\text{M} \rangle $ by Alice, with $\hat{O}_\text{A}\in \mathfrak{A}_\text{M}^\text{NW} (\mathcal{V}_\text{A})$, may influence the measurement of $\hat{O}_\text{B}$ by Rob.

In Ref.~\cite{PhysRevA.107.L030203}, we showed an explicit example in which such a nonlocal effect occurs. In particular, we considered a Minkowski single particle localized in the left wedge with respect to the Newton-Wigner scheme and we demonstrated that its preparation affects measurements of Rindler observables in the right wedge.

It has been argued that selective operations cannot be used to instantly send information to other disjoint regions \cite{Redhead1995-REDMAA-2, CLIFTON20011, VALENTE2014147, RevModPhys.90.045003}. On the other hand, nonselective local preparations of states in $\mathcal{O}_\text{A}$ affecting local measurements of observables in $\mathcal{O}_\text{B}$ lead to violations of causality. In the ARbM scenario, the independence between local preparations of states and local measurements of observables by the two experimenters is not guaranteed by the unitarity of $\hat{O}_\text{A}$. Hence, the nonlocal effect predicted here is in contradiction with causality. In principle, Alice can unitarily prepare a local state to instantly send information to Rob if they are localized with respect to the Newton-Wigner scheme of their respective frame.

We highlight the novelty of this subsection by recalling the features of the Newton-Wigner localization in QFT. It has been pointed out that noncovariance under Lorentz transformation and superluminal spreading of the wave function are unavoidable features of the Newton-Wigner scheme \cite{RevModPhys.21.400, PhysRevD.10.3320, PhysRevD.22.377}. However, the theory is not problematic at all if one fixes the Lorentz frame and only considers the spacelike hypersurface $t = 0$. We remark that the acausal effect of the Hegerfeldt theorem can only be detected at times different from $t = 0$, though arbitrary close from it. Here, instead, we only consider the initial hypersurface $t = 0$ and no Lorentz boost, since at that time the accelerated observer has zero velocity in the inertial frame. Instead, we consider noninertial coordinate transformations. We show the noncovariance under diffeomorphism and we derive the violation of causality occurring at the instant $t = 0$. Hence, we conclude that the Newton-Wigner scheme lack GR covariance in addition to special relativistic covariance and that acausal effects already occur at $t = 0$ for an inertial and a noninertial observer.

\section{AQFT scheme}\label{AQFT_localization_scheme_in_curved_spacetime}

Hereafter, we consider a massive scalar real fields in 3+1 dimensional flat spacetime. We refer to $ \hat{\phi} (t,\vec{x})$ as the representation of the field in the Minkowski frame $(t,\vec{x}) = (t,x,y,z)$, whereas $\hat{\Phi}_\nu (T,\vec{X})$ is the field in the $\nu$-Rindler frame with coordinates $(T,\vec{X}) = (T,X,Y,Z)$. The identity relating $ \hat{\phi} (t,\vec{x})$ to $\hat{\Phi}_\nu (T,\vec{X})$ is
\begin{equation}\label{scalar_transformation_Rindler}
\hat{\Phi}_\nu (T,\vec{X}) = \hat{\phi} (t_\nu(T,\vec{X}), \vec{x}_\nu(T,\vec{X})),
\end{equation}
where
\begin{subequations}\label{Rindler_coordinate_transformation}
\begin{align}
& t_\nu(T,\vec{X}) = \frac{e^{s_\nu a Z}}{c a} \sinh(c a T), \quad x_\nu(T,\vec{X}) = X, \\
 & y_\nu(T,\vec{X}) = Y,  \quad z_\nu(T,\vec{X}) = s_\nu \frac{e^{s_\nu a Z}}{a} \cosh(c a T)
\end{align}
\end{subequations}
is the coordinate transformation from one frame to the other. Equation (\ref{scalar_transformation_Rindler}) reflects the fact that the field transforms as a scalar under diffeomorphisms.

In this section, we study the AQFT localization scheme for both inertial and accelerated observers. We consider the algebraic approach to QFT, which is based on the notion of local algebra $\mathfrak{A}(\mathcal{E})$ associated to each spacetime event $\mathcal{E}$. $\mathfrak{A}(\mathcal{E})$ is defined as the algebra generated by the field operator $\hat{\phi}(x^\mu)$, with $x^\mu$ as the Minkowski coordinate representing $\mathcal{E}$. As a consequence of Eq.~(\ref{scalar_transformation_Rindler}), the local algebra $\mathfrak{A}(\mathcal{E})$ can be equivalently defined with respect to the field operator $\hat{\Phi}_\nu (T,\vec{X})$ if the wedge $\nu$ and the Rindler coordinate $X^\mu$ represent the same event $\mathcal{E}$. Crucially, we find that local field operators in each frame (i.e., $\hat{\phi}(x^\mu)$ and $\hat{\Phi}_\nu(X^\mu)$) emerge as a coordinatization to the same local algebraic structure given by $\mathfrak{A}(\mathcal{E})$ and, hence, they provide different representations of the same theory. This gives an unifying formulation of the AQFT for both inertial and accelerated frame, which reflects the frame independent notion of spacetime events $\mathcal{E}$.

The AQFT localization scheme is based on the definition of the local algebras $\mathfrak{A}(\mathcal{E})$. The operator $\hat{O}$ is said to be localized in the spacetime event $\mathcal{E}$ with respect to the AQFT scheme if it is an element of $\mathfrak{A}(\mathcal{E})$. We also say that the state $| \psi \rangle$ is localized in $\mathcal{E}$ over the background $| \Omega \rangle $ if it is the result of local operations on $| \Omega \rangle $ in $\mathcal{E}$; explicitly, this means that there is an $\hat{O} \in \mathfrak{A}(\mathcal{E})$ such that $| \psi \rangle = \hat{O} | \Omega \rangle$. In addition to spacetime events $\mathcal{E}$, one can consider extended regions $\mathcal{O}$ and say that the observable $\hat{O}$ and the state $| \psi \rangle = \hat{O} | \Omega \rangle$ are localized in $\mathcal{O}$ over the background $| \Omega \rangle$ if $\hat{O}$ belongs to the local algebra $\mathfrak{A}(\mathcal{O})$.

To see an example, consider the Rindler spacetime coordinate $(T,\vec{X})$ and the real scalar field $\hat{O} = \hat{\Phi}_\nu(T,\vec{X})$ evaluated at the point $(T,\vec{X})$ and at the wedge $\nu$. The observable $\hat{\Phi}_\nu(T,\vec{X})$ is an element of $\mathfrak{A}(\mathcal{E})$, where $\mathcal{E}$ is the spacetime event represented by $(T,\vec{X})$ and $\nu$ in the Rindler coordinate system. Hence, we say that the observable $\hat{O}$ is localized in $\mathcal{E}$ with respect to the AQFT scheme. Also, the state $| \psi \rangle = \hat{O} | \Omega \rangle$ is localized in $\mathcal{E}$ over the background $| \Omega \rangle$.

Notice that, as a consequence of Eq.~(\ref{scalar_transformation_Rindler}), the operator $\hat{O} = \hat{\Phi}_\nu(T,\vec{X})$ can be written in terms of the Minkowski scalar field as $\hat{\phi} (t_\nu(T,\vec{X}), \vec{x}_\nu(T,\vec{X}))$. The Minkowski coordinate point $(t,\vec{x}) = (t_\nu(T,\vec{X}), \vec{x}_\nu(T,\vec{X}))$ represents the same event $\mathcal{E}$ identified by the Rindler coordinate $(T,\vec{X})$ and the wedge $\nu$. We find that if an observable is localized in an event $\mathcal{E}$ with respect to an accelerated observer, it is also localized in $\mathcal{E}$ with respect to an inertial observer. Hence, the notion of localization in AQFT appears to be frame independent, in the sense that inertial and accelerated observers agree on the region in which states and observables are localized. As a result, we obtain a notion of localization that obeys physical invariance under diffeomorphisms, in agreement with the GR theory.

The frame independent notion of the AQFT scheme can be described in terms of local preparation of states over the background $| \Omega \rangle$ and local measurements of observables. We consider an inertial and an accelerated experimenter, Alice and Rachel, that prepare states in each of their frames. If they are both localized in the same spacetime region, they share the same algebra of operators and may, in principle, prepare the same state. The only difference is given by the respective coordinate systems by means of which they describe the spacetime. Analogously, an inertial and an accelerated experimenter, Bob and Rob, can perform the same measurement if they are localized in the same region of spacetime.

In addition to the invariance under diffeomorphisms, the relativistic notion of causality is included in the formulation of the AQFT scheme. This is a consequence of the canonical communication relations
\begin{equation}\label{psi_commutation}
\left[ \hat{\phi}(t,\vec{x}), \hat{\phi}(t',\vec{x}') \right] = i \hbar \Delta_\text{KG}(t-t', \vec{x} - \vec{x}')
\end{equation}
satisfied by the bosonic field $ \hat{\phi} (t,\vec{x})$. Here,
\begin{align}\label{PauliJordan_function}
\Delta_\text{KG}(t,\vec{x}) = & \frac{- i}{(2 \pi)^3} \int_{\mathbb{R}^3} \frac{d^3 k}{2 \omega(\vec{k})} \nonumber \\
& \times \left[ e^{-i\omega(\vec{k})t + i\vec{k} \cdot \vec{x}} - e^{i\omega(\vec{k})t - i\vec{k} \cdot \vec{x}} \right]
\end{align}
is the Pauli–Jordan function (i.e., retarded minus advanced propagator) and
\begin{equation} \label{omega_k}
\omega(\vec{k}) = \sqrt{\left(\frac{mc^2}{\hbar}\right)^2 + c^2 |\vec{k}|^2}
\end{equation}
is the frequency of the particle with momentum $\vec{k}$ and mass $m$. Equations (\ref{psi_commutation}) and (\ref{PauliJordan_function}) imply that field operators commute if they are spacelike separated. Hence, for any couple of operators $\hat{O}_\text{A} \in \mathfrak{A}(\mathcal{O}_\text{A})$ and $\hat{O}_\text{B} \in \mathfrak{A}(\mathcal{O}_\text{B})$, we find that $[\hat{O}_\text{A}, \hat{O}_\text{B}] = 0$ if $\mathcal{O}_\text{A}$ and $\mathcal{O}_\text{B}$ are spacelike separated regions of spacetime. This is known as the microcausality axiom of AQFT \cite{haag1992local}.

The commutativity between the local algebras $\mathfrak{A}(\mathcal{O}_\text{A})$ and $\mathfrak{A}(\mathcal{O}_\text{B})$ ensures that measurements conducted in $\mathcal{O}_\text{A}$ and $\mathcal{O}_\text{B}$ do not influence each other. Also, it guarantees the independence between unitary local preparation of states in $\mathcal{O}_\text{A}$ and local measurements in $\mathcal{O}_\text{B}$. To see this, assume that the state $| \psi \rangle = \hat{O}_\text{A} | \Omega \rangle$ is prepared by means of the unitary local operator $\hat{O}_\text{A} \in \mathfrak{A}(\mathcal{O}_\text{A})$ and consider any observables $\hat{O}_\text{B} \in \mathfrak{A}(\mathcal{O}_\text{B})$. By using the commutation identity $[\hat{O}_\text{A}, \hat{O}_\text{B}] = 0$ the unitary condition $\hat{O}_\text{A}^\dagger \hat{O}_\text{A} = 1$, prove that
\begin{equation}\label{local_unitary_operator_measurament_A_omega}
\langle \Omega | \hat{O}_\text{A}^\dagger \hat{O}_\text{B} \hat{O}_\text{A} | \Omega \rangle = \langle \Omega | \hat{O}_\text{A}^\dagger \hat{O}_\text{A} \hat{O}_\text{B} | \Omega \rangle = \langle \Omega | \hat{O}_\text{B} | \Omega \rangle,
\end{equation}
which implies that $| \psi \rangle$ is strictly localized in $\mathcal{O}_\text{A}$ [Eq.~(\ref{KnightLicht_property_Omega})].

This result can be extended to the case of nonselective operations by means of local Kraus operators $\hat{K}_i \in \mathfrak{A}(\mathcal{O}_\text{A})$ satisfying the condition $\sum_i \hat{K}_i^\dagger \hat{K}_i = 1$. The action of $\hat{K}_i$ on the background $|\Omega \rangle$ gives the state
\begin{equation}\label{rho_Kraus}
\hat{\rho} = \frac{\sum_i \hat{K}_i |\Omega \rangle \langle \Omega | \hat{K}_i^\dagger}{\sum_i \langle \Omega | \hat{K}_i^\dagger \hat{K}_i |\Omega \rangle},
\end{equation}
which appears to be strictly localized, since it satisfies the identity
\begin{equation}\label{Tr_rho_A_nonselective}
\text{Tr} (\hat{\rho}\hat{O}_\text{B}) = \langle \Omega| \hat{O}_\text{B} |\Omega \rangle.
\end{equation}

In the AQFT scheme, the strict localization property is only guaranteed for nonselective operations. Selective preparations of states, instead, do not satisfy the strict localization property and, hence, they violate the independence between the operations in $\mathcal{O}_\text{A}$ and $\mathcal{O}_\text{B}$ even when they are spacelike separated. Such a result is independent of the nature of the experimenters in $\mathcal{O}_\text{A}$ and $\mathcal{O}_\text{B}$. They can be both inertial, both accelerated or one inertial and the other one accelerated; in all of these cases, nonlocal effects occur if states are prepared via selective operations. This is due to the fact that inertial and accelerated observers agree on the localization of states and observables. We find that, in the AQFT scheme, all the proposed scenarios (i.e., ABM, RaRbR, RaRbM and ARbM) are guaranteed to satisfy the strict localization property only for nonselective preparations of states.

\section{Modal scheme}\label{Modal_localization_scheme_Rindler}

In this section, we consider the scalar field in the Minkowski ($ \hat{\phi} (t,\vec{x})$) and the Rindler ($\hat{\Phi}_\nu (T,\vec{X})$) spacetime and we study the modal scheme in each frame. Such a localization scheme is based on the representation of single particle states as positive frequency solutions of the respective Klein-Gordon equation.

This section is organized as follows. In Sec.~\ref{Modal_representation}, we give a brief introduction on the modal representation of Minkowski and Rindler particles in terms of their modal wave functions; a more detailed analysis can be found in Ref.~\cite{PhysRevD.107.045012}. In Sec.~\ref{Modal_localization}, we use the modal representation of particles to define localized states and observables from the point of view of inertial and accelerated observers. In Sec.~\ref{Modal_acausal_notcovatian}, we show that the modal scheme is incompatible with relativistic causality and physical invariance under diffeomorphism and, hence, it does not entail any genuine notion of localization in the QFTCS regime, at variance with the AQFT scheme.

\subsection{Modal representation of Minkowski and Rindler particles}\label{Modal_representation}

The Minkowski-Klein-Gordon equation is
\begin{equation} \label{Klein_Gordon}
\left[ \eta^{\mu\nu} \partial_\mu \partial_\nu - \left( \frac{mc}{\hbar} \right)^2 \right] \hat{\phi}(t,\vec{x}) = 0,
\end{equation}
where $\eta^{\mu\nu} = \text{diag}(-c^{-2},1,1,1)$ is the Minkowski metric. The Rindler-Klein-Gordon equation, instead, is
\begin{align}\label{Rindler_Klein_Gordon}
& \left\lbrace  c^2 e^{s_\nu 2 a Z} \left[ \partial_1^2 + \partial_2^2 - \left( \frac{mc}{\hbar} \right)^2 \right] \right. \nonumber \\
& \left. - \partial_0^2 + c^2 \partial_3^2 \right\rbrace \hat{\Phi}_\nu (T,\vec{X})= 0.
\end{align}

The respective positive frequency solutions of each field equation are
\begin{equation}\label{free_modes}
f(\vec{k},t,\vec{x}) =  \sqrt{\frac{\hbar}{(2\pi)^3 2 \omega(\vec{k})}} e^{-i\omega(\vec{k})t + i\vec{k} \cdot \vec{x}}
\end{equation}
for the Minkowski-Klein-Gordon equation (\ref{Klein_Gordon}) and
\begin{equation}\label{F_Rindler}
F_\nu(\Omega,\vec{K}_\perp,T,\vec{X}) = \tilde{F}(\Omega,\vec{K}_\perp,s_\nu Z) e^{ i \vec{K}_\perp \cdot \vec{X}_\perp - i \Omega T },
\end{equation}
with
\begin{align} \label{F_tilde_Rindler}
\tilde{F}(\Omega,\vec{K}_\perp,Z) = & \frac{1}{2 \pi^2 c} \sqrt{ \frac{\hbar}{a} \left| \sinh \left( \frac{\pi \Omega}{c a} \right) \right| } \nonumber \\
& \times K_{i \Omega /ca} \left( \kappa (\vec{K}_\perp) \frac{e^{aZ}}{a} \right),
\end{align}
for the Rindler-Klein-Gordon equation (\ref{Rindler_Klein_Gordon}) \cite{RevModPhys.80.787}. The functions appearing in Eq.~(\ref{F_tilde_Rindler}) are
\begin{equation}\label{kappa_k_perp}
\kappa (\vec{K}_\perp) = \sqrt{ \left( \frac{m c}{\hbar} \right)^2 + |\vec{K}_\perp|^2 }
\end{equation}
and $K_\zeta (\xi)$ as the modified Bessel function of the second kind. For simplicity, we define $\vec{\theta} = (\Omega, \vec{K}_\perp)$ as the collection of Rindler quantum numbers.

The modes of each frame are orthogonal with respect to the Klein-Gordon product
\begin{align}\label{KG_scalar_product}
( \psi, \psi' )_\text{KG} = & \frac{i}{\hbar} \int_{\mathbb{R}^3} d^3x \left[ \psi^*(t,\vec{x}) \partial_0  \psi'(t,\vec{x}) \right. \nonumber \\
& \left. -  \psi'(t,\vec{x}) \partial_0 \psi^*(t,\vec{x}) \right].
\end{align}
Explicitly, the orthogonality condition for $f(\vec{k},t,\vec{x})$ is
\begin{subequations}\label{KG_scalar_product_orthonormality_f}
\begin{align}
& ( f(\vec{k}), f(\vec{k}') )_\text{KG} = \delta^3(\vec{k}-\vec{k}'), \\
& ( f^*(\vec{k}), f^*(\vec{k}') )_\text{KG} = -\delta^3(\vec{k}-\vec{k}'),\\
& ( f(\vec{k}), f^*(\vec{k}') )_\text{KG} = 0,
\end{align}
\end{subequations}
while for $F_\nu(\vec{\theta},T,\vec{X})$,
\begin{subequations}\label{KG_scalar_curved_product_orthonormality_F}
\begin{align}
& ( F(\vec{\theta}), F(\vec{\theta}'))_\text{KG} = \delta^3 (\vec{\theta}-\vec{\theta}') , \\
 & ( F^*(\vec{\theta}), F^*(\vec{\theta}'))_\text{KG}  = - \delta^3 (\vec{\theta}-\vec{\theta}') ,\\
 & ( F(\vec{\theta}), F^*(\vec{\theta}'))_\text{KG} = 0.
\end{align}
\end{subequations}
 
By means of the modes $f(\vec{k},t,\vec{x})$ and $F_\nu(\vec{\theta},T,\vec{X})$, the fields $ \hat{\phi} (t,\vec{x})$ and $\hat{\Phi}_\nu (T,\vec{X})$ can be decomposed as
\begin{equation} \label{free_field}
\hat{\phi}(t,\vec{x}) = \int_{\mathbb{R}^3} d^3 k \left[ f(\vec{k},t,\vec{x}) \hat{a}(\vec{k}) + f^*(\vec{k},t,\vec{x}) \hat{a}^\dagger(\vec{k}) \right],
\end{equation}
and
\begin{align} \label{Rindler_scalar_decomposition}
& \hat{\Phi}_\nu (T,\vec{X}) \nonumber \\
 = & \int_0^{\infty} d\Omega \int_{\mathbb{R}^2} d^2 K_\perp \left[ F_\nu(\Omega,\vec{K}_\perp,T, \vec{X})\hat{A}_\nu(\Omega,\vec{K}_\perp)  \right. \nonumber \\
& \left. + F_\nu^*(\Omega,\vec{K}_\perp,T, \vec{X})\hat{A}_\nu^\dagger(\Omega,\vec{K}_\perp) \right],
\end{align}
respectively. Here, $\hat{a}(\vec{k})$ is the Minkowski annihilation operator of the particles with momentum $\vec{k}$ satisfying the canonical commutation relation
\begin{align}\label{Minkowski_canonical_commutation}
& [\hat{a}(\vec{k}),\hat{a}^\dagger(\vec{k}')]  = \delta^3(\vec{k}-\vec{k}'),
& [\hat{a}(\vec{k}),\hat{a}(\vec{k}')] = 0.
\end{align}
Conversely, $\hat{A}_\nu(\Omega,\vec{K}_\perp)$ is the annihilator of the $\nu$-Rindler particle with frequency $\Omega$ and transverse momentum $\vec{K}_\perp$ satisfying
\begin{align}\label{Rindler_canonical_commutation}
& [\hat{A}_\nu(\vec{\theta}),\hat{A}_{\nu'}^\dagger(\vec{\theta}')]  = \delta_{\nu\nu'} \delta^3(\vec{\theta}-\vec{\theta}'),
& [\hat{A}_\nu(\vec{\theta}),\hat{A}_{\nu'}(\vec{\theta}')] = 0.
\end{align}

In the modal representation, single particle states are described by the positive frequency modes $f(\vec{k},t,\vec{x})$ and $F_\nu(\Omega,\vec{K}_\perp,T,\vec{X})$. The function $f(\vec{k},t,\vec{x})$ represents the Minkowski particle with momentum $\vec{k}$ evolved in time with respect to the Schrödinger picture; whereas $F_\nu(\Omega,\vec{K}_\perp,T,\vec{X})$ represents the $\nu$-Rindler particle with frequency $\Omega$ and transverse momentum $\vec{K}_\perp$.

For any Minkowski-Fock state
\begin{equation}\label{free_state_decomposition}
| \psi \rangle  = \left[ \sum_{n=1}^\infty \frac{1}{\sqrt{n!}} \int_{\mathbb{R}^{3n}} d^{3n} \textbf{k}_n \tilde{\psi}_n (\textbf{k}_n) \prod_{l=1}^n \hat{a}^\dagger(\vec{k}_l)  + \tilde{\psi}_0 \right] | 0_\text{M} \rangle,
\end{equation}
its representative in the modal representation is
\begin{align} \label{free_wave_function}
\psi_n (t, \textbf{x}_n) = & \left( \frac{2 m c^2}{\hbar^2} \right)^{n/2} \int_{\mathbb{R}^{3n}} d^{3n} \textbf{k}_n  \nonumber \\
& \times \tilde{\psi}_n (\textbf{k}_n) \prod_{l=1}^n f(\vec{k}_l, t, \vec{x}_l).
\end{align}
Conversely, the Rindler-Fock state
\begin{align} \label{general_Fock_expansion_Rindler}
| \Psi \rangle  = & \left[ \sum_{n=1}^\infty \frac{1}{\sqrt{n!}} \sum_{\bm{\nu}_n} \int_{[(0,\infty)\otimes\mathbb{R}^2]^n} d^{3n}\bm{\theta}_n  \right. \nonumber \\
& \left. \times \tilde{\Psi}_n (\bm{\nu}_n,\bm{\theta}_n) \prod_{l=1}^n \hat{A}_{\nu_l}^\dagger(\vec{\theta}_l) +  \tilde{\Psi}_0 \right] | 0_\text{L}, 0_\text{R} \rangle,
\end{align}
is represented by
\begin{align}\label{wavefunction_F}
\Psi_n (T, \bm{\nu}_n, \textbf{X}_n) = & \left( \frac{2 m c^2}{\hbar^2} \right)^{n/2} \int_{[(0,\infty)\otimes\mathbb{R}^2]^n} d^{3n}\bm{\theta}_n \nonumber \\
& \times \tilde{\Psi}_n (\bm{\nu}_n,\bm{\theta}_n)    \prod_{l=1}^n F_{\nu_l}(\vec{\theta}_l,T,\vec{X}_l).
\end{align}
Here, the vectors $\textbf{k}_n = (\vec{k}_1, \dots, \vec{k}_n)$ and $(\bm{\nu}_n,\bm{\theta}_n) = ((\nu_1,\vec{\theta}_1), \dots, (\nu_n , \vec{\theta}_n))$ are collections of quantum numbers; $\textbf{x}_n = (\vec{x}_1, \dots, \vec{x}_n)$ and $\textbf{X}_n = (\vec{X}_1, \dots, \vec{X}_n)$ are $3n$ vectors collecting $n$ position coordinates in the respective frame; the sum $\sum_{\bm{\nu}_n}$ runs over all possible combinations of $\nu_l \in \{ \text{L}, \text{R} \}$. When $n=0$, we assume $\psi_0 = \tilde{\psi}_0$ and $\Psi_0 = \tilde{\Psi}_0$. Hereafter, we refer to $\tilde{\psi}_n (\textbf{k}_n)$ and $\psi_n (t, \textbf{x}_n) $ as modal wave functions of $| \psi \rangle$ in, respectively, the momentum and the position space; analogously, $\tilde{\Psi}_n (\bm{\nu}_n,\bm{\theta}_n)$ and $\Psi_n (T, \bm{\nu}_n, \textbf{X}_n)$ are modal wave functions of the Rindler-Fock state $| \Psi \rangle$.

\subsection{Modal localization of states and observables}\label{Modal_localization}

The modal representation of states by means of the wave functions $\psi_n (0, \textbf{x}_n) $ and $\Psi_n (0, \bm{\nu}_n, \textbf{X}_n)$ naturally lead to a notion of localization for Minkowski and Rindler particles based on the support of their modal wave functions in position space.  We say that the Minkowski-Fock state $| \psi \rangle \in \mathcal{H}_\text{M}$ is localized in the Minkowski coordinate region $\mathcal{V}$ at $t=0$ over the Minkowski vacuum $| 0_\text{M} \rangle$ and with respect to the Minkowski modal scheme if its wave function $\psi_n (0, \textbf{x}_n)$ is supported in $\mathcal{V}^n$, in the sense that $\psi_n (0, \textbf{x}_n) = 0$ when there is an $l \in \{ 1, \dots, n \}$ such that $\vec{x}_l \notin \mathcal{V}$. Analogously the Rindler-Fock state $| \Psi \rangle \in \mathcal{H}_{\text{L},\text{R}}$ is localized in the wedge $\nu$ and in the volume $\mathcal{V}$ at $T=0$ over the Rindler vacuum $| 0_\text{L}, 0_\text{R} \rangle$ and with respect to the Rindler modal scheme if $\Psi_n (0, \bm{\nu}_n, \textbf{X}_n)$ is supported in $\nu^n \otimes \mathcal{V}^n$, i.e., $\Psi_n (0, \bm{\nu}_n, \textbf{X}_n) = 0$ when there is an $l \in \{ 1, \dots, n \}$ such that $\nu_l \neq \nu$ or $\vec{x}_l \notin \mathcal{V}$.

The localization of Minkowski-Fock states over the Minkowski vacuum $| 0_\text{M} \rangle$ and Rindler-Fock states over the Rindler vacuum $| 0_\text{L}, 0_\text{R} \rangle$ provide a natural definition of local operators. We say that $\hat{O}$ is localized in the Minkowski coordinate region $\mathcal{V}$ with respect to the Minkowski modal scheme if the state $\hat{O} | 0_\text{M} \rangle$ is localized in $\mathcal{V}$ over the Minkowski vacuum $| 0_\text{M} \rangle$ with respect to the Minkowski modal scheme. Conversely, $\hat{O}$ is localized in the wedge $\nu$ and in the volume $\mathcal{V}$ with respect to the Rindler modal scheme if $\hat{O} | 0_\text{L}, 0_\text{R} \rangle $ is localized in $\nu$ and $\mathcal{V}$ over the Rindler vacuum $| 0_\text{L}, 0_\text{R} \rangle$ with respect to the Rindler modal scheme. This naturally leads to the definition of local algebras $\mathfrak{A}_\text{M}^\text{mod}(\mathcal{V})$ and $\mathfrak{A}_\nu^\text{mod}(\mathcal{V})$ respectively generated by operators localized in $\mathcal{V}$ with respect to the Minkowski modal scheme and by operators localized in $\nu$ and $\mathcal{V}$ with respect to the Rindler modal scheme. We can also consider the limiting situation in which the region $\mathcal{V}$ is point-like and define the modal algebras $\mathfrak{A}_\text{M}^\text{mod}(\vec{x})$ and $\mathfrak{A}_\nu^\text{mod}(\vec{X})$ in the Minkowski coordinate $\vec{x}$ and in the $\nu$-Rindler coordinate $\vec{X}$.

The generators of the algebras $\mathfrak{A}_\text{M}^\text{mod}(\mathcal{V})$ and $\mathfrak{A}_\text{M}^\text{mod}(\vec{x})$ can be found by inverting Eq.~(\ref{free_wave_function}) and by writing Eq.(\ref{free_state_decomposition}) in terms of the wave function $\psi_n (0, \textbf{x}_n)$. The inverse of Eq.~(\ref{free_wave_function}) is
\begin{align} \label{free_wave_function_inverse}
\tilde{\psi}_n (\textbf{k}_n) = & \left[ \frac{\hbar}{(2 \pi)^3 m c^2} \right]^{n/2} \int_{\mathbb{R}^{3n}} d^{3n} \textbf{x}_n \nonumber\\
& \times  \psi_n (0, \textbf{x}_n) \prod_{l=1}^n \left[ \sqrt{\omega(\vec{k}_l)}  e^{- i \vec{k}_l \cdot \vec{x}_l} \right],
\end{align}
which can be plugged in Eq.~(\ref{free_state_decomposition}) to obtain
\begin{equation}\label{free_state_decomposition_antiparticles}
| \psi \rangle  = \hat{a}_\text{mod}^\dagger[\psi] | 0_\text{M} \rangle,
\end{equation}
with
\begin{equation}\label{c_psi_space}
\hat{a}_\text{mod}^\dagger[\psi]  =  \sum_{n=0}^\infty \frac{1}{\sqrt{n!}}   \int_{\mathbb{R}^{3n}} d^{3n} \textbf{x}_n  \psi_n (0, \textbf{x}_n)  \prod_{l=1}^n \hat{a}_\text{mod}^\dagger (\vec{x}_l),
\end{equation}
and with
\begin{equation}\label{a_mod_a}
\hat{a}_\text{mod} (\vec{x})  = \int_{\mathbb{R}^3} d^3 k  \sqrt{\frac{\hbar \omega(\vec{k})}{(2 \pi)^3 mc^2}}  e^{i \vec{k} \cdot \vec{x}} \hat{a}(\vec{k}) .
\end{equation}
By definition, $| \psi \rangle $ is localized in $\mathcal{V}$ over $| 0_\text{M} \rangle$ with respect to the Minkowski modal scheme if $\Psi_n (0, \bm{\nu}_n, \textbf{X}_n) $ is supported in $\mathcal{V}^n$. This means that the operator $\hat{a}_\text{mod}^\dagger[\psi] $ is an element of $ \mathfrak{A}_\text{M}^\text{mod}(\mathcal{V})$, with $\mathcal{V}$ as the support of $\psi_n (0, \textbf{x}_n)$. Also, we find that the generators of the algebra $ \mathfrak{A}_\text{M}^\text{mod}(\mathcal{V})$ are the operators $\hat{a}_\text{mod} (\vec{x})$ with $\vec{x} \in \mathcal{V}$; whereas, $\hat{a}_\text{mod} (\vec{x})$ is the generator of $\mathfrak{A}_\text{M}^\text{mod}(\vec{x})$.

The inverse of Eq.~(\ref{wavefunction_F}) is
\begin{align}\label{wavefunction_F_inverse}
\tilde{\Psi}_n (\bm{\nu}_n,\bm{\theta}_n)  = & \left( \frac{2 m c^2}{\hbar^2} \right)^{n/2}  \int_{\mathbb{R}^{3n}} d^{3n} \textbf{X}_n  \nonumber \\
& \times \Psi_n (0, \bm{\nu}_n, \textbf{X}_n) \prod_{l=1}^n  \mathcal{F}_{\nu_l}^*(\vec{\theta}_l,0,\vec{X}_l),
\end{align}
with
\begin{equation}\label{FF_nu_F_nu}
\mathcal{F}_\nu (\Omega, \vec{K}_\perp, T ,\vec{X}) = \frac{\hbar \Omega}{m c^2} F_\nu (\Omega, \vec{K}_\perp, T ,\vec{X}).
\end{equation}
Equation (\ref{wavefunction_F_inverse}) is a result of
\begin{equation}\label{FF_nu_F_nu_delta}
\int_{\mathbb{R}^3} d^3X  \frac{2 m c^2}{ \hbar^2 } F_\nu(\vec{\theta},0,\vec{X})  \mathcal{F}_\nu^*(\vec{\theta}',0,\vec{X}) = \delta^3(\vec{\theta}-\vec{\theta}'),
\end{equation}
which can be derived from Eqs.~(\ref{F_Rindler}), (\ref{FF_nu_F_nu}) and from
\begin{align}\label{alpha_tilde_approx_3_inverse}
& \int_0^\infty dz  \frac{2 \Omega}{\hbar a z}  \tilde{F}(\Omega, \vec{K}_\perp,Z_\text{R}(z)) \tilde{F}(\Omega', \vec{K}_\perp,Z_\text{R}(z)) \nonumber \\
 =& \frac{1}{4 \pi^2} \delta(\Omega-\Omega').
\end{align}
The proof for Eq.~(\ref{alpha_tilde_approx_3_inverse}) is shown in Appendix \ref{Proof_of_alpha_tilde_approx_3_inverse}.

By using Eq.~(\ref{wavefunction_F_inverse}) in Eq.~(\ref{general_Fock_expansion_Rindler}) we obtain
\begin{equation}
| \Psi \rangle  = \hat{A}_{\text{mod},\nu}^\dagger[\Psi] | 0_\text{L}, 0_\text{R} \rangle,
\end{equation}
with
\begin{align}\label{A_mod_Psi_n}
\hat{A}_{\text{mod},\nu}^\dagger[\Psi]  = & \sum_{n=0}^\infty \frac{1}{\sqrt{n!}} \sum_{\bm{\nu}_n}   \int_{\mathbb{R}^{3n}} d^{3n} \textbf{X}_n  \nonumber \\
& \times \Psi_n (0, \bm{\nu}_n, \textbf{X}_n) \prod_{l=1}^n \hat{A}_{\text{mod},\nu_l}^\dagger (\vec{X}_l) ,
\end{align}
and with
\begin{equation}\label{a_mod_a_nu}
\hat{A}_{\text{mod},\nu} (\vec{X})  = \int_{\theta_1>0} d^3 \theta \frac{\sqrt{2 m c^2}}{\hbar} \mathcal{F}_\nu(\vec{\theta},0,\vec{X}) \hat{A}_\nu(\vec{\theta}).
\end{equation}
By using the definition of localized states with respect to the Rindler modal scheme, we find that $\mathfrak{A}_\nu^\text{mod}(\mathcal{V})$ is generated by the operators $\hat{A}_{\text{mod},\nu} (\vec{X})$ with $\vec{X} \in \mathcal{V}$. For fixed Rindler coordinate $\vec{X}$, the operator $\hat{A}_{\text{mod},\nu} (\vec{X})$ generates the algebra $\mathfrak{A}_\nu^\text{mod}(\vec{X})$.

By means of the local algebras $\mathfrak{A}_\text{M}^\text{mod}(\vec{x})$, it is possible to extend the definition of localized states with respect to the Minkowski modal scheme to include any background $| \Omega \rangle$ that differs from the Minkowski vacuum $| 0_\text{M} \rangle$. In particular, we say that the state $| \psi \rangle = \hat{O} | \Omega \rangle$ is localized in $\vec{x}$ over the background $| \Omega \rangle$ with respect to the Minkowski modal scheme if $\hat{O}$ is an element of $\mathfrak{A}_\text{M}^\text{mod}(\vec{x})$. Analogously, we say that the state $| \Psi \rangle = \hat{O} | \Omega \rangle$ is localized in $\nu$ and $\vec{X}$ over the background $| \Omega \rangle$ with respect to the Rindler modal scheme if $\hat{O}$ is an element of $\mathfrak{A}_\nu^\text{mod}(\vec{X})$.

\subsection{The modal scheme is noncovariant and acausal}\label{Modal_acausal_notcovatian}

The noncovariant behavior of the modal scheme can be proven by noticing that the Rindler modal scheme is not compatible with the Minkowski modal scheme, in the sense that no state or operator is localized in the same spacetime event with respect to both schemes. Mathematically, this means that the algebras $\mathfrak{A}_\text{M}^\text{mod}(\vec{x})$ and $\mathfrak{A}_\nu^\text{mod}(\vec{X})$ do not coincide, even if $\vec{x}$ and $\vec{X}$ represent the same point in the $\nu$ wedge, i.e., $\vec{x} = \vec{x}_\nu(0,\vec{X})$. This is a consequence of the nontrivial Bogoliubov transformation relating the Rindler modal operators $\hat{A}_{\text{mod},\nu} (\vec{X}) $ [Eq.~(\ref{a_mod_a_nu})] to their Minkowski counterpart $\hat{a}_\text{mod} (\vec{x})$ [Eq.~(\ref{a_mod_a})], i.e.,
\begin{align}\label{a_mod_M_A_mod_nu}
& \hat{a}_\text{mod} (\vec{x}) \nonumber \\
 = & \sum_{\nu=\{\text{L},\text{R}\}} \int_{\mathbb{R}^3} d^3X  \left[ f_{\text{mod},\nu \mapsto \text{mod}}^+ (\vec{x},\vec{X})  \hat{A}_{\text{mod},\nu} (\vec{X}) \right. \nonumber \\
& \left. + f_{\text{mod},\nu \mapsto \text{mod}}^- (\vec{x},\vec{X}) \hat{A}_{\text{mod},\nu}^\dagger (\vec{X}) \right],
\end{align}
with
\begin{align}
f_{\text{mod},\nu \mapsto \text{mod}}^\pm (\vec{x},\vec{X}) = & \int_{\mathbb{R}^3} d^3 k  \int_{\theta_1>0} d^3 \theta \sqrt{\frac{ \omega(\vec{k})}{4 \pi^3 \hbar}}  e^{i \vec{k} \cdot \vec{x}} \nonumber \\
& \times \alpha_\nu(\vec{k},\pm \vec{\theta})   F_\nu^*(\pm \vec{\theta},0,\vec{X}).
\end{align}

Equation (\ref{a_mod_M_A_mod_nu}) can be derived from the Bogoliubov transformation relating Minkowski operators to Rindler operators
\begin{align}\label{Rindler_Bogoliubov_transformations}
\hat{a}(\vec{k}) =  &  \sum_{\nu=\{\text{L},\text{R}\}}\int_{\theta_1>0} d^3 \theta \left[ \alpha_\nu(\vec{k},\vec{\theta})  \hat{A}_\nu(\vec{\theta}) \right.\nonumber \\
& \left. + \alpha_\nu(\vec{k},-\vec{\theta}) \hat{A}^\dagger_\nu(\vec{\theta}) \right],
\end{align}
with
\begin{align} \label{alpha}
\alpha_\nu(\vec{k},\vec{\theta}) = & \int_{\mathbb{R}^3} d^3 x \frac{\theta(s_\nu z)}{\hbar}  \left[  \frac{s_\nu \theta_1}{a z} + \omega(\vec{k}) \right] f^*( \vec{k}, 0,  \vec{x} ) \nonumber \\
& \times \tilde{F}(\vec{\theta},  s_\nu Z_\nu(z)) e^{ i \vec{\theta}_\perp \cdot \vec{x}_\perp},
\end{align}
and from the inverse of (\ref{a_mod_a_nu}), i.e.,
\begin{equation}\label{a_mod_a_nu_inverse}
\hat{A}_\nu(\vec{\theta}) = \int_{\mathbb{R}^3} d^3X  \frac{\sqrt{2 m c^2}}{ \hbar } F_\nu^*(\vec{\theta},0,\vec{X}) \hat{A}_{\text{mod},\nu} (\vec{X}).
\end{equation}
Equation (\ref{Rindler_Bogoliubov_transformations}) can be derived from Eqs.~(\ref{scalar_transformation_Rindler}) and (\ref{Rindler_scalar_decomposition}) and by inverting (\ref{free_field}) by means of Eq.~(\ref{KG_scalar_product_orthonormality_f}). Equation (\ref{a_mod_a_nu_inverse}), instead, can be proven by using Eq.~(\ref{FF_nu_F_nu_delta}).

Equation (\ref{a_mod_M_A_mod_nu}) implies that $\mathfrak{A}_\text{M}^\text{mod}(\vec{x}) \neq \mathfrak{A}_\nu^\text{mod}(\vec{X})$ when $\vec{x} = \vec{x}_\nu(0,\vec{X})$. As a result, we find that the modal localization scheme does not satisfy GR covariance under diffeomorphism and cannot be used to describe local physical phenomena in the QFTCS regime.

To prove the incompatibility between the modal scheme and the relativistic causality, one can notice that operators localized in different regions of the space do not generally commute. In the case of Minkowski modal scheme, one can see this by computing
\begin{subequations}\label{a_mod_commutation}
\begin{align}
& [\hat{a}_\text{mod}(\vec{x}), \hat{a}_\text{mod}^\dagger(\vec{x}') ] = \int_{\mathbb{R}^3} d^3 k  \frac{\hbar \omega(\vec{k})}{ mc^2}  \frac{e^{i \vec{k} \cdot (\vec{x} - \vec{x}')}}{(2 \pi)^3} , \label{a_mod_commutation_a}\\
& [\hat{a}_\text{mod}(\vec{x}), \hat{a}_\text{mod}(\vec{x}') ] = 0
\end{align}
\end{subequations}
from Eqs.~(\ref{Minkowski_canonical_commutation}) and (\ref{a_mod_a}). By using  Eqs.~(\ref{Rindler_canonical_commutation}) and (\ref{a_mod_a_nu}), instead, one can derive the commutation relation
\begin{subequations}\label{A_mod_commutation}
\begin{align}
[ \hat{A}_{\text{mod},\nu} (\vec{X}), \hat{A}_{\text{mod},\nu'}^\dagger (\vec{X}')] = & \delta_{\nu\nu'} \int_{\theta_1>0} d^3 \theta \frac{2 m c^2}{\hbar^2} \nonumber \\
& \times \mathcal{F}_\nu(\vec{\theta},0,\vec{X}) \mathcal{F}_\nu^*(\vec{\theta},0,\vec{X}'), \\
  [ \hat{A}_{\text{mod},\nu} (\vec{X}), \hat{A}_{\text{mod},\nu'} (\vec{X}')]=&0,
\end{align}
\end{subequations}
which implies that the operators $\hat{A}_{\text{mod},\nu} (\vec{X}) $ and $\hat{A}_{\text{mod},\nu}^\dagger (\vec{X}') $ do not commute with each other, even if $\vec{X}$ is different from $\vec{X}'$. Also, by means of Eqs.~(\ref{a_mod_M_A_mod_nu}) and (\ref{A_mod_commutation}) it can be proven that $\hat{a}_\text{mod} (\vec{x}) $ does not commute with $\hat{A}_{\text{mod},\nu} (\vec{X}) $ nor $\hat{A}_{\text{mod},\nu}^\dagger (\vec{X}) $, even if the Minkowski coordinate $\vec{x}$ represents the same event of the $\nu$-Rindler coordinate $\vec{X}$.

As a result, we find that operators localized in different regions with respect to the Minkowski modal scheme or with respect to the Rindler modal scheme are not guaranteed to commute. Explicitly, this means that $[\hat{O}_\text{A}, \hat{O}_\text{B}] \neq 0$ for some couples of operators $\hat{O}_\text{A} $ and $\hat{O}_\text{B} $ which are localized in disjoint regions with respect to any modal scheme, i.e., $\hat{O}_\text{A} \in \mathfrak{A}_\text{M}^\text{mod}(\mathcal{V}_\text{A})$ and $\hat{O}_\text{B} \in \mathfrak{A}_\text{M}^\text{mod}(\mathcal{V}_\text{B})$ with $\mathcal{V}_\text{A} \cap \mathcal{V}_\text{B} \neq \varnothing$, or $\hat{O}_\text{A} \in \mathfrak{A}_\nu^\text{mod}(\mathcal{V}_\text{A})$ and $\hat{O}_\text{B} \in \mathfrak{A}_\nu^\text{mod}(\mathcal{V}_\text{B})$ with $\mathcal{V}_\text{A} \cap \mathcal{V}_\text{B} \neq \varnothing$, or $\hat{O}_\text{A} \in \mathfrak{A}_\text{M}^\text{mod}(\mathcal{V}_\text{A})$ and $\hat{O}_\text{B} \in \mathfrak{A}_\nu^\text{mod}(\mathcal{V}_\text{B})$ with $\mathcal{V}_\text{A}$ and $\mathcal{V}_\text{B}$ representing disjoint regions of the spacetime.

The commutativity of operators localized in spacelike separated regions is a necessary condition for the compatibility between the localization scheme and the relativistic causality. When two operators $\hat{O}_\text{A}$ and $\hat{O}_\text{B}$ do not commute, the corresponding measurements are not independent. Hence, the local measurements conducted by an observer in the space region $\mathcal{V}_\text{A}$ may influence the results of measurements carried out in a disjoint region $\mathcal{V}_\text{B}$.

Also, the preparations of the state $| \psi \rangle$ in $\mathcal{V}_\text{A}$ may influence measurements in $\mathcal{V}_\text{B}$, even if $| \psi \rangle$ is prepared via unitary operation. This result may be interpreted in terms of the modal ABM, RaRbR, RaRbM and ARbM scenarios, in which each experimenter prepares local states and carries out local measurements using the respective modal scheme. Specifically, in the modal ABM scheme, both Alice and Bob describe localized states and observables with respect to the Minkowski modal scheme over the Minkowski vacuum $| 0_\text{M} \rangle$; in the modal RaRbR and RaRbM scenarios, the experimenters use the Rindler modal scheme over, respectively, the Rindler vacuum $|0_\text{L},0_\text{R} \rangle$ and the the Minkowski vacuum $| 0_\text{M} \rangle$; in the modal ARbM scenario Alice uses the Minkowski modal scheme over $| 0_\text{M} \rangle$, while Rob uses the Rindler modal scheme. In all of these cases, the preparation of states by the experimenter A may influence measurements carried out by the experimenter B.

Crucially, the violation of independence between preparations of states and measurements of observables in disjoint regions also occurs for unitary operations. In other words, the strict localization property (\ref{KnightLicht_property_Omega}) is not guaranteed for unitary prepared states that are localized with respect to the modal scheme. To see this, assume that the state $| \psi \rangle = \hat{O}_\text{A} | \Omega \rangle$ is localized in $\mathcal{O}_\text{A}$ over $| \Omega \rangle$, with $\hat{O}_\text{A}$ satisfying the unitary condition $\hat{O}_\text{A}^\dagger \hat{O}_\text{A} = 1$, and consider any operator $\hat{O}_\text{B}$ measured by the experimenter B in $\mathcal{O}_\text{B}$. By using the noncommutative identity for operators localized in different regions, we find that
\begin{align}
& \langle \psi | \hat{O}_\text{B} | \psi \rangle = \langle \Omega | \hat{O}_\text{A}^\dagger \hat{O}_\text{B} \hat{O}_\text{A} | \Omega \rangle \nonumber \\
\neq & \langle \Omega | \hat{O}_\text{A}^\dagger \hat{O}_\text{A}  \hat{O}_\text{B}| \Omega \rangle  = \langle \Omega |  \hat{O}_\text{B}| \Omega \rangle,
\end{align}
which means that the unitary preparation of $| \psi \rangle$ by experimenter A is able to influence measurements by experimenter B in $\hat{O}_\text{B}$. This leads to a violation of causality.

An additional acausal effect is given by the instantaneous spreading of modal wave functions $\psi_n (t, \textbf{x}_n)$ and $\Psi_n (T, \bm{\nu}_n, \textbf{X}_n)$. It can be shown that if a wave function has compact support in $\mathcal{V}$ at initial time $t=T=0$, then, immediately after, it develops tails outside the light-cone of $\mathcal{V}$. This issue is due to the fact that $\psi_n (t, \textbf{x}_n)$ and $\Psi_n (T, \bm{\nu}_n, \textbf{X}_n)$ are linear combinations of products of, respectively, the modes $f(\vec{k},t,\vec{x})$ and $F_\nu (\Omega, \vec{K}_\perp, T ,\vec{X})$. It can be proven that if a positive frequency mode has compact support in $\mathcal{V}$, then its time derivative cannot be supported in $\mathcal{V}$ \cite{Afanasev:1996nm}. This means that $ \partial_0 \psi_n (t, \textbf{x}_n) |_{t=0}$ and $\psi_n (0, \textbf{x}_n)$ cannot be simultaneously compactly supported; the same holds for $ \partial_0 \Psi_n (T, \bm{\nu}_n, \textbf{X}_n) |_{T=0}$ and $\Psi_n (0, \bm{\nu}_n, \textbf{X}_n)$.

The modal scheme does not entail any genuine notion of localization in QFTCS since it violates relativistic causality and covariance under diffeomorphism. Only the AQFT framework is able to provide a valid description for special relativistic and GR local phenomena. In Sec.~\ref{The_modal_scheme_converges_to_the_AQFT_scheme}, we will show that the modal scheme converges to the AQFT scheme in the nonrelativistic limit. Hence, we will find that the modal scheme is suited for the description of local phenomena only in the nonrelativistic regime.

\section{Comparison between localization schemes}\label{Comparison_between_localization_schemes_Rindler}

\begin{table}
\begin{center}
\centering
\begin{tabular}{| >{\raggedright\arraybackslash}m{11em} || >{\centering\arraybackslash}m{7em} |  >{\centering\arraybackslash}m{3.5em} |}
\hline
 & AQFT scheme & modal scheme\\
\hline
\hline
GR covariance and causality hold  & Yes & No \\
\hline
Operators in disjoint spatial regions commute  & Yes & No \\
\hline
The strict localization property [Eq.~(\ref{KnightLicht_property_Omega})] at $t=T=0$ is guaranteed & Only for local nonselective preparations & No \\
\hline
\end{tabular}
\end{center}
\caption{Summary table of the differences between the AQFT and the modal localization schemes in the Minkowski and the Rindler frame.}\label{Comparison_between_localization_schemes_QFTCS_Table} 
\end{table}

In this section, we compare the AQFT and the modal localization schemes presented in Secs.~\ref{AQFT_localization_scheme_in_curved_spacetime} and \ref{Modal_localization_scheme_Rindler}, respectively. The results are summarized in Table \ref{Comparison_between_localization_schemes_QFTCS_Table}.

In Sec.~\ref{AQFT_localization_scheme_in_curved_spacetime}, by showing the compatibility between the Minkowski and the Rindler AQFT schemes, we proved that the GR covariance holds. The modal scheme, instead, lacks this property as the local algebras $\mathfrak{A}_\text{M}^\text{mod}(\vec{x})$ and $\mathfrak{A}_\nu^\text{mod}(\vec{X})$ are not equivalent, even if $\vec{x}$ and $\vec{X}$ represent the same event in the $\nu$ wedge, i.e., $\vec{x} = \vec{x}_\nu(0,\vec{X})$. Also, the relativistic causality in the AQFT schemes is ensured by the microcausality axiom (i.e., the commutativity condition for field operators in spacelike separated regions) and by the fact that nonselectively prepared local states are always strictly localized. In the modal scheme, instead, wave functions instantly propagate and the strict localization property is not guaranteed even for unitarily prepared states, due to the noncommutativity relation between operators localized in disjoint regions. For these reasons we conclude that, between the two schemes, only the AQFT scheme gives a genuine description of local phenomena.

The possibility to compare the AQFT and the modal scheme (i.e., $\mathfrak{A}(\mathcal{O})$ and $\mathfrak{A}_\text{M}^\text{mod}(\mathcal{V})$ for inertial experimenters, $\mathfrak{A}(\mathcal{O})$ and $\mathfrak{A}_\nu^\text{mod}(\mathcal{V})$ for accelerated experimenters) is hindered by the fact that the regions $\mathcal{O}$ and $\mathcal{V}$ are conceptually different, as $\mathcal{O}$ refers to a spacetime region, whereas $\mathcal{V}$ is a Minkowski or Rindler coordinate region. A direct comparison can only be obtained by defining the local algebras in spatial regions with respect to the AQFT scheme. By considering the Cauchy hypersurface $t=T=0$ and the dynamics induced by the hyperbolic differential equations (\ref{Klein_Gordon}) and (\ref{Rindler_Klein_Gordon}), we know that the global algebra is generated by $\hat{\phi}(0,\vec{x})$ and its conjugate $\hat{\pi}(0,\vec{x}) =  - \partial_0 \hat{\phi}(t,\vec{x})|_{t=0}$, or, equivalently, by $\hat{\Phi}_\nu(0,\vec{X})$ and $\hat{\Pi}_\nu(0,\vec{X}) =  - \partial_0 \hat{\Phi}_\nu(T,\vec{X})|_{T=0}$. This fact naturally leads to the definition of the algebra $\mathfrak{A}_\text{M}^\text{AQFT}(\mathcal{V})$ generated by $\hat{\phi}(0,\vec{x})$ and $\hat{\pi}(0,\vec{x})$ with $\vec{x} \in \mathcal{V}$ and the algebra $\mathfrak{A}_\nu^\text{AQFT}(\mathcal{V})$ generated by $\hat{\Phi}_\nu(0,\vec{X})$ and $\hat{\Pi}_\nu(0,\vec{X}) $ with $\vec{X} \in \mathcal{V}$. The AQFT and the modal scheme can be directly compared by means of the algebras $\mathfrak{A}_\text{M}^\text{AQFT}(\mathcal{V})$ and $\mathfrak{A}_\text{M}^\text{mod}(\mathcal{V})$, or, equivalently, by means of $\mathfrak{A}_\nu^\text{AQFT}(\mathcal{V})$ and $\mathfrak{A}_\nu^\text{mod}(\mathcal{V})$.

We also define the local algebras in space points with respect to the AQFT scheme $\mathfrak{A}_\text{M}^\text{AQFT}(\vec{x})$ and $\mathfrak{A}_\nu^\text{AQFT}(\vec{X})$ as the one generated by the couple $\hat{\phi}(0,\vec{x})$ and $\hat{\pi}(0,\vec{x})$ and the couple $\hat{\Phi}_\nu(0,\vec{X})$ and $\hat{\Pi}_\nu(0,\vec{X}) $, respectively. Notice that the local algebra in the Rindler frame $\mathfrak{A}_\nu^\text{AQFT}(\vec{X})$ is equivalent to the algebra $\mathfrak{A}_\text{M}^\text{AQFT}(\vec{x})$ in the Minkowski frame if $\vec{x} = \vec{x}_\nu(0, \vec{X})$. This can be proven by deriving the one-to-one map 
\begin{subequations}
\begin{align}
& \hat{\Phi}_\nu (0,\vec{X}) = \hat{\phi} (0, \vec{x}_\nu(0,\vec{X})), \\
& \hat{\Pi}_\nu(0,\vec{X})  = s_\nu a z_\nu(Z) \hat{\pi} (0, \vec{x}_\nu(0,\vec{X})),
\end{align}
\end{subequations}
which can be obtained by means of Eq.~(\ref{scalar_transformation_Rindler}) and the chain rule
\begin{equation}\label{chain_rule_Rindler}
\frac{\partial}{\partial T} = s_\nu az \frac{\partial}{\partial t}  + s_\nu ac^2t  \frac{\partial}{\partial z}.
\end{equation}

The incompatibility between the Minkowski modal and the AQFT scheme has been explicitly shown in Ref.~\cite{localization_QFT} by deriving the Bogoliubov transformation
\begin{align}\label{a_mod_psi_pi}
\hat{a}_\text{mod} (\vec{x}) & = \int_{\mathbb{R}^3} d^3 x' \left[ f_{\hat{\phi}\mapsto \text{mod}}(\vec{x} - \vec{x}') \hat{\phi}(0, \vec{x}') \right. \nonumber\\
& \left. + f_{\hat{\pi}\mapsto \text{mod}}(\vec{x} - \vec{x}') \hat{\pi}(0, \vec{x}')  \right],
\end{align}
with
\begin{subequations}
\begin{align}
& f_{\hat{\phi}\mapsto \text{mod}}(\vec{x}) = \int_{\mathbb{R}^3} d^3 k  \frac{\omega (\vec{k}) e^{i \vec{k} \cdot \vec{x}}}{ (2 \pi)^3 \sqrt{2 m c^2}},\\
 & f_{\hat{\pi}\mapsto \text{mod}}(\vec{x}) = \frac{-i}{\sqrt{2 m c^2}} \delta^3(\vec{x}).
\end{align}
\end{subequations}
The inequality $\mathfrak{A}_\text{M}^\text{mod}(\vec{x}) \neq \mathfrak{A}_\text{M}^\text{AQFT}(\vec{x})$ is a result of the fact that $f_{\hat{\phi}\mapsto \text{mod}}(\vec{x})$ has support in the entire space $\mathbb{R}^3$.

Analogously, the incompatibility between the Rindler modal and the AQFT scheme can be proved by showing that operators that are localized with respect to the one of the two schemes are not localized with respect to the other. To see this, consider a Rindler single particle creator, defined as
\begin{equation}\label{A_mod_Psi_1}
\hat{A}_{\text{mod},\nu}^\dagger [\Psi] = \int_{\theta_1>0} \tilde{\Psi} (\vec{\theta})  \hat{A}_\nu^\dagger(\vec{\theta}).
\end{equation}
By means of Eq.~(\ref{a_mod_a_nu_inverse}), one can also consider
\begin{equation}\label{A_mod_Psi}
\hat{A}_{\text{mod},\nu}^\dagger [\Psi] = \int_{\mathbb{R}^3} d^3X \Psi (\vec{X})  \hat{A}_{\text{mod},\nu}^\dagger(\vec{X}),
\end{equation}
with $\Psi (\vec{X}) = \Psi (0, \vec{X})$ and
\begin{equation}
\Psi (T, \vec{X}) =  \frac{\sqrt{2 m c^2}}{\hbar}  \tilde{\Psi} (\vec{\theta})  F_\nu(\vec{\theta},T,\vec{X})
\end{equation}
as the wave function in position space. Equation (\ref{A_mod_Psi}) is the restriction of Eq.~(\ref{A_mod_Psi_n}) to the $\nu$-Rindler single particle space, whose elements are characterized by vanishing wave functions $\Psi_n (0, \bm{\nu}_n, \textbf{X}_n)$ except for $n=1$ and for $\nu_1 = \nu$; $\Psi (T, \vec{X})$ is, then, defined to be equal to $ \Psi_1 (T, \nu, \vec{X})$. The operator $\hat{A}_{\text{mod},\nu}^\dagger [\Psi]$ is localized in the wedge $\nu$ and in the region $\mathcal{V}$ with respect to the Rindler modal scheme (i.e., $\hat{A}_{\text{mod},\nu}^\dagger [\Psi] \in \mathfrak{A}_\nu^\text{mod}(\mathcal{V})$) if $\Psi (\vec{X})$  is supported in $\mathcal{V}$.

Consider an operator $\hat{A}_{\text{mod},\nu}^\dagger [\Psi]$ such that $\tilde{\Psi} (\vec{\theta})$ is real. By using the orthonormality condition (\ref{KG_scalar_curved_product_orthonormality_F}) in Eq.~(\ref{Rindler_scalar_decomposition}), we can write $\hat{A}_{\text{mod},\nu}^\dagger [\Psi]$ in terms of the Rindler fields $\hat{\Phi}_\nu(0,\vec{X}), \hat{\Pi}_\nu(0,\vec{x}) $ as
\begin{align}\label{A_Psi_1_2}
& \hat{A}_{\text{mod},\nu}^\dagger [\Psi]\nonumber \\
 = & -\int_{\theta_1>0} \tilde{\Psi} (\vec{\theta}) (\hat{\Phi}_\nu, F_\nu^*(\vec{\theta}))_\text{KG} \nonumber\\
= & - \frac{\hbar}{\sqrt{2 m c^2}} (  \hat{\Phi}_\nu, \Psi^* )_\text{KG}\nonumber\\
= & - \frac{i}{\sqrt{2 m c^2}} \int_{\mathbb{R}^3} d^3X \left[ \hat{\Phi}_\nu(0,\vec{X}) \left. \partial_0 \Psi^*(T,\vec{X})\right|_{T=0} \right. \nonumber \\
& \left. + c^2 \Psi^*(\vec{X})  \hat{\Pi}_\nu(0,\vec{X}) \right].
\end{align}
If $\Psi (\vec{X})$ is supported in $\mathcal{V}$, the operator $\hat{A}_\text{mod}^\dagger [\Psi]$ is localized with respect to the Rindler modal scheme. However, it is not localized with respect to the AQFT scheme due to $\partial_0 \Psi(T,\vec{X})|_{T=0}$ acting as a smearing function for $\hat{\Phi}_\nu(0,\vec{X})$ in Eq.~(\ref{A_Psi_1_2}). The function $\Psi(T,\vec{X})$ is a positive frequency solution of the field equation; therefore, if it is supported in $\mathcal{V}$, its time derivative $\partial_0 \Psi(T,\vec{X})$ is not.

In Sec.~\ref{The_modal_scheme_converges_to_the_AQFT_scheme}, we will show how the two localization schemes converge in the nonrelativistic limit. This will imply that the modal operators $\hat{A}_{\text{mod},\nu}^\dagger [\Psi] \in \mathfrak{A}_\nu^\text{mod} (\mathcal{V})$ can be approximated by some local field operators $\hat{A}_{\text{AQFT},\nu}^\dagger [\Psi] \in \mathfrak{A}_\nu^\text{AQFT} (\mathcal{V})$. The intuition is that if $\Psi(\vec{X})$ is supported in $\mathcal{V}$, then its time derivative $\partial_0 \Psi(T,\vec{X}) |_{T=0}$ is approximately vanishing outside $\mathcal{V}$. Consequently, the right hand side of Eq.~(\ref{A_Psi_1_2}) is made of field operators approximately smeared over $\mathcal{V}$ and, hence, the single particle modal operator $\hat{A}_\text{mod}^\dagger [\Psi]$ is approximately local in $\mathcal{V}$ with respect to the AQFT scheme.

\section{Localization in the nonrelativistic regime}\label{Localization_in_NRQFTCS}

In the previous sections, we considered different localization schemes in the Minkowski and the Rindler frames over a general background state $| \Omega \rangle$. In Sec.~\ref{Comparison_between_localization_schemes_Rindler}, we showed that the AQFT and the modal schemes are incompatible.

In this section, we consider the nonrelativistic limit of QFT and the NonRelativistic limit of Quantum Field Theory in Curved Spacetime (NRQFTCS) \cite{PhysRevD.107.045012}. We show that the two schemes converge in each frame and we detail the features of the resulting nonrelativistic localization schemes.

In Ref.~\cite{localization_QFT}, we discussed the nonrelativistic limit of the AQFT and the modal schemes in Minkowski spacetime over the Minkowski vacuum $| 0_\text{M} \rangle$. We showed the convergence between the two schemes and the emergence of the familiar Born localization, which is based on the representation of states by means of wave functions whose modulo square gives the probability to find the state in each space point. States that are localized in space regions $\mathcal{V}$ appear indistinguishable from $| 0_\text{M} \rangle$ outside $\mathcal{V}$. As a result, we found that, in the nonrelativistic ABM scenario, the Reeh-Schlieder nonlocality \cite{haag1992local, Redhead1995-REDMAA-2, PhysRevA.58.135, Reeh:1961ujh} is suppressed, which means that the independence between preparations of states and measurements in disjoint regions of space is recovered in the nonrelativistic regime.

Here, we extend these results by studying the nonrelativistic localization scheme in the Rindler frame over a general background state $| \Omega \rangle$. We consider the nonrelativistic limit of Rindler scalar fields \cite{PhysRevD.107.045012} and we prove the convergence between the AQFT and the modal schemes. Then we detail the features of this unified localization scheme in the NRQFTCS regime.

We show that if $| \Omega \rangle = |0_\text{L},0_\text{R} \rangle$, the modal wave functions acquire the Born probabilistic notion of finding particles in each point and the indistinguishably from the Rindler vacuum $ |0_\text{L},0_\text{R} \rangle$ outside their support. This is a consequence of the convergence to the AQFT scheme---which entails a genuine notion of locality---and the factorization of the Rindler nonrelativistic global Hilbert space and the Rindler vacuum into, respectively, Rindler nonrelativistic local Fock spaces and local vacua. Due to the emergence of the Born localization scheme in the Rindler frame, we find that, in the nonrelativistic RaRbR scenario, the preparation of nonrelativistic states by an accelerated experimenter over the Rindler vacuum $|0_\text{L},0_\text{R} \rangle$ does not affect nonrelativistic measurements carried out by another accelerated observer in a separated region.

Conversely, if the background $| \Omega \rangle$ is not the Rindler vacuum $|0_\text{L},0_\text{R} \rangle$, the independence between preparation and measurement in disjoint regions does not generally hold. In particular, in the RaRbM scenario, the background Minkowski vacuum $|0_\text{M} \rangle$ is entangled between the Rindler nonrelativistic local Fock spaces and, hence, it produces nonlocal effects that are similar to the Reeh-Schlieder nonlocality.

Finally, in the ARbM scenario, the two experimenters (Alice and Rob) do not share the same notion of nonrelativistic limit \cite{PhysRevD.107.085016} and, hence, their nonrelativistic localization schemes are incompatible. This leads to a persistence of the Reeh-Schlieder nonlocal effect in nonrelativistic ARbM experiments.

The section is organized as follows. In Sec~\ref{The_modal_scheme_converges_to_the_AQFT_scheme}, we prove the convergence between the modal and the AQFT scheme in the nonrelativistic limit. In Sec.~\ref{Born_scheme_in_NRQFTCS} we discuss the emergence of the Born localization scheme in the nonrelativistic limit of each frame; the results will be interpreted in terms of the nonrelativistic ABM and RaRbR scenarios. In Secs.~\ref{The_modal_scheme_over_general_background_does_not_converge_to_the_Born_scheme} and \ref{AliceRob_nonlocality_is_not_suppressed_by_the_nonrelativistic_limit}, we respectively study the RaRbM and the ARbM scenarios, with particular focus on the strict localization property.

\subsection{Convergence between the modal scheme and the AQFT scheme}\label{The_modal_scheme_converges_to_the_AQFT_scheme}

In Ref.~\cite{localization_QFT}, we demonstrated the convergence between the Minkowski modal and the AQFT scheme in the nonrelativistic limit. We considered the nonrelativistic condition in Minkowski spacetime, defined by
\begin{equation}\label{non_relativistic_limit}
\left| \frac{\hbar \omega(\vec{k})}{mc^2} - 1 \right| \lesssim \epsilon,
\end{equation}
where $\epsilon \ll 1$ is the nonrelativistic parameter representing the maximum ratio between the nonrelativistic energy $E=\hbar \omega - mc^2$ and the mass energy $mc^2$. The state $| \psi \rangle $ is said to be nonrelativistic if the corresponding modal wave function in momentum space $\tilde{\psi}_n (\textbf{k}_n)$ is such that
\begin{align}\label{non_relativistic_limit_wavefunction}
& \tilde{\psi}_n (\textbf{k}_n) \approx 0 \text{ if there is an } l \in \{ 1, \dots , n \} \nonumber \\
&  \text{such that }\left| \frac{\hbar \omega(\vec{k}_l)}{mc^2} - 1 \right| \gg \epsilon.
\end{align}
By following Ref.~\cite{Papageorgiou_2019}, we defined the bandlimited subspace $\mathcal{H}_\text{M}^\epsilon$ as the Fock space of Minkowski particles satisfying Eq.~(\ref{non_relativistic_limit_wavefunction}). By restricting Eq.~(\ref{a_mod_psi_pi}) to $\mathcal{H}_\text{M}^\epsilon$, we found that $\left. \hat{a}_\text{mod}(\vec{x}) \right|_{\mathcal{H}_\text{M}^\epsilon} \approx \left. \hat{a}_\text{AQFT}(\vec{x}) \right|_{\mathcal{H}_\text{M}^\epsilon}$, with
\begin{equation}\label{alpha_tilde_psi_Pi}
\hat{a}_\text{AQFT}(\vec{x}) = \sqrt{\frac{m c^2}{2 \hbar^2}} \hat{\phi}(0,\vec{x}) - \frac{i}{\sqrt{2 m c^2}} \hat{\pi}(0,\vec{x}),
\end{equation}
as an element of $\mathfrak{A}_\text{M}^\text{AQFT}(\vec{x})$. This proves the convergence between the Minkowski modal and the AQFT scheme in the nonrelativistic limit (\ref{non_relativistic_limit}).

Here, we extend these results by considering the Rindler frame and we show the convergence between the Rindler modal and the AQFT scheme in the nonrelativistic limit. By following Ref.~\cite{PhysRevD.107.045012}, we define the nonrelativistic condition in Rindler spacetime as
\begin{equation}\label{non_relativistic_limit_curved}
\left| \frac{\hbar \Omega}{mc^2} - 1 \right| \lesssim \epsilon,
\end{equation}
with $\epsilon \ll 1$ as the nonrelativistic parameter. We say that $| \Psi \rangle $ is nonrelativistic if the corresponding wave function $\tilde{\Psi}_n (\bm{\nu}_n,\bm{\theta}_n)$ is non-vanishing only for frequencies $\vec{\theta}$ satisfying Eq.~(\ref{non_relativistic_limit_curved}). Explicitly, this means that
\begin{align}\label{non_relativistic_limit_curved_wavefunction}
& \tilde{\Psi}_n (\bm{\nu}_n,\bm{\theta}_n) \approx 0 \text{ if there is an } l \in \{ 1, \dots , n \} \nonumber \\
&   \text{such that }\left| \frac{\hbar \Omega(\vec{\theta}_l)}{mc^2} - 1 \right| \gg \epsilon,
\end{align}
where $\Omega(\vec{\theta}) = \theta_1$. In this subsection, we prove that operators that are localized with respect to the Rindler modal scheme are also localized with respect to the AQFT scheme.

We firstly adopt a method similar to the one used in Ref.~\cite{Papageorgiou_2019}. In particular we define the space of Rindler nonrelativistic particles $\mathcal{H}_{\text{L},\text{R}}^\epsilon$ as the subspace of the Rindler-Fock space $\mathcal{H}_{\text{L},\text{R}}$ that contains particles with frequencies satisfying the nonrelativistic condition (\ref{non_relativistic_limit_curved}). Also, we define the operator
\begin{equation}\label{alpha_tilde_Psi_Pi}
\hat{A}_{\text{AQFT},\nu}(\vec{X}) = \sqrt{\frac{m c^2}{2 \hbar^2}} \hat{\Phi}_\nu(0,\vec{X}) - \frac{i}{\sqrt{2 m c^2}} \hat{\Pi}_\nu(0,\vec{X}),
\end{equation}
which generates the local field algebra $\mathfrak{A}_\nu^\text{AQFT}(\vec{X})$.

The right hand side of Eq.~(\ref{alpha_tilde_Psi_Pi}) can be written in terms of Rindler-Fock operators as
\begin{align}\label{alpha_tilde_Psi_Pi_2}
& \hat{A}_{\text{AQFT},\nu}(\vec{X}) \nonumber \\
 = & \int_0^{\infty} d\Omega \int_{\mathbb{R}^2} d^2 K_\perp \left[ f_{\hat{A} \mapsto \text{AQFT},\nu}(\vec{X},\Omega,\vec{K}_\perp) \hat{A}_\nu(\Omega,\vec{K}_\perp)  \right. \nonumber \\
& \left. + f_{\hat{A}^\dagger \mapsto \text{AQFT},\nu}(\vec{X},\Omega,\vec{K}_\perp) \hat{A}_\nu^\dagger(\Omega,\vec{K}_\perp) \right],
\end{align}
with
\begin{subequations}\label{f_A_alpha_tilde}
\begin{align}
f_{\hat{A} \mapsto \text{AQFT},\nu}(\vec{X},\Omega,\vec{K}_\perp) = & \left( \sqrt{\frac{m c^2}{2 \hbar^2}} + \frac{\Omega}{\sqrt{2 m c^2}} \right)  \nonumber \\
& \times F_\nu(\Omega,\vec{K}_\perp,0, \vec{X}) , \\
f_{\hat{A}^\dagger \mapsto \text{AQFT},\nu}(\vec{X},\Omega,\vec{K}_\perp) =  & \left( \sqrt{\frac{m c^2}{2 \hbar^2}} - \frac{\Omega}{\sqrt{2 m c^2}} \right)  \nonumber \\
& \times F_\nu^*(\Omega,\vec{K}_\perp,0, \vec{X}).
\end{align}
\end{subequations}
Notice that when $\Omega$ satisfies the nonrelativistic condition (\ref{non_relativistic_limit_curved}), Eq.~(\ref{f_A_alpha_tilde}) can be approximated by $f_{\hat{A} \mapsto \text{AQFT},\nu}(\vec{X},\Omega,\vec{K}_\perp) \approx (\sqrt{2m c^2}/ \hbar) \mathcal{F}_\nu(\Omega,\vec{K}_\perp,0,\vec{X}) $ and $f_{\hat{A}^\dagger \mapsto \text{AQFT},\nu}(\vec{X},\Omega,\vec{K}_\perp) \approx 0$. Hence, by restricting the operators $\hat{A}_{\text{AQFT},\nu}(\vec{X})$ and $\hat{A}_{\text{mod},\nu} (\vec{X})$ to $\mathcal{H}_{\text{L},\text{R}}^\epsilon$, we find that
\begin{equation}\label{A_AQFT_A_mod_nonrelativistic}
\left. \hat{A}_{\text{AQFT},\nu}(\vec{X}) \right|_{\mathcal{H}_{\text{L},\text{R}}^\epsilon} \approx \left. \hat{A}_{\text{mod},\nu} (\vec{X}) \right|_{\mathcal{H}_{\text{L},\text{R}}^\epsilon},
\end{equation}
which means that any element of the modal algebra $\mathfrak{A}_\nu^\text{mod}(\vec{X})$ can be approximated to an element of $\mathfrak{A}_\nu^\text{AQFT}(\vec{X})$ when restricted to $\mathcal{H}_{\text{L},\text{R}}^\epsilon$.

The Hilbert space $\mathcal{H}_{\text{L},\text{R}}^\epsilon $ describes nonrelativistic Rindler particles that are prepared over the Rindler vacuum $| 0_\text{L}, 0_\text{R} \rangle$. Hence, the results of Eq.~(\ref{A_AQFT_A_mod_nonrelativistic}) can be used in the nonrelativistic RaRbR scenario, where all the experimenters are accelerated and the background state $| \Omega \rangle$ is precisely the Rindler vacuum $| 0_\text{L}, 0_\text{R} \rangle$. However, we also want to consider scenarios in which $| \Omega \rangle$ is different from $| 0_\text{L}, 0_\text{R} \rangle$ (e.g., RaRbM scenario) or where one of the two observers uses nonrelativistic Minkowski operators to prepare states (e.g., ARbM scenario). To include all of these cases we now adopt an algebraic approach to the nonrelativistic limit. 

Starting from the AQFT scheme, we introduce the local algebra $\mathfrak{A}_\nu^{\text{AQFT},\epsilon} (\mathcal{V}) $ as the subset of $\mathfrak{A}_\nu^\text{AQFT} (\mathcal{V}) $ that contains combinations of products of Rindler operators $\hat{A}_\nu(\Omega,\vec{K}_\perp)$ satisfying the nonrelativistic condition (\ref{non_relativistic_limit_curved}). By definition, the algebra $\mathfrak{A}_\nu^{\text{AQFT},\epsilon} (\mathcal{V}) $ is generated by operators of the form of
\begin{equation}\label{A_AQFT_Psi_1}
\hat{A}_{\text{AQFT},\nu}^\dagger [\Psi_\epsilon] = \int_{\mathbb{R}^3} d^3X \Psi_\epsilon (\vec{X})  \hat{A}_{\text{AQFT},\nu}^\dagger(\vec{X}),
\end{equation}
where $\Psi_\epsilon (\vec{X}) $ is a function supported in $\mathcal{V}$ that satisfies the following nonrelativistic condition
\begin{subequations}\label{nonrelativistic_condition_AQFT}
\begin{align}
& \int_{\mathbb{R}^3} d^3X f_{\hat{A} \mapsto \text{AQFT},\nu}^*(\vec{X},\Omega,\vec{K}_\perp) \Psi_\epsilon (\vec{X}) \approx 0  \nonumber \\
&  \text{if } \left|  \frac{\hbar \Omega}{m c^2} - 1 \right| \gg \epsilon, \\
& \int_{\mathbb{R}^3} d^3X f_{\hat{A}^\dagger \mapsto \text{AQFT},\nu}^*(\vec{X},\Omega,\vec{K}_\perp) \Psi_\epsilon (\vec{X})  \approx 0  \nonumber \\
&  \text{if } \left|  \frac{\hbar \Omega}{m c^2} - 1 \right| \gg \epsilon.
\end{align}
\end{subequations}
Equation (\ref{nonrelativistic_condition_AQFT}) can be obtained by plugging Eq.~(\ref{alpha_tilde_Psi_Pi_2}) in Eq.~(\ref{A_AQFT_Psi_1}) and by assuming that $\hat{A}_{\text{AQFT},\nu}^\dagger [\Psi_\epsilon]$ is a linear combination of Rindler operators $\hat{A}_\nu(\Omega,\vec{K}_\perp)$ satisfying the nonrelativistic condition (\ref{non_relativistic_limit_curved}).

Analogously, we define the local algebra $\mathfrak{A}_\nu^{\text{mod},\epsilon} (\mathcal{V}) $ as the subset of $\mathfrak{A}_\nu^\text{mod} (\mathcal{V}) $ that contains combinations of products of Rindler operators $\hat{A}_\nu(\Omega,\vec{K}_\perp)$ satisfying the nonrelativistic condition (\ref{non_relativistic_limit_curved}). Explicitly, this means that $\mathfrak{A}_\nu^{\text{mod},\epsilon} (\mathcal{V}) $ is generated by the operators $\hat{A}_{\text{mod},\nu}^\dagger [\Psi_\epsilon]$ [Eq.~(\ref{A_mod_Psi})] with $\Psi_\epsilon (\vec{X})$ supported in $\mathcal{V}$ and satisfying
\begin{equation}\label{nonrelativistic_condition_mod}
\int_{\mathbb{R}^3} d^3X \mathcal{F}_\nu^*(\Omega,\vec{K}_\perp,0,\vec{X}) \Psi_\epsilon (\vec{X})  \approx 0 \text{ if } \left|  \frac{\hbar \Omega}{m c^2} - 1 \right| \gg \epsilon.
\end{equation}
Equation (\ref{nonrelativistic_condition_mod}) is a consequence of Eq.~(\ref{A_mod_Psi_1}) and the nonrelativistic condition for $\hat{A}_{\text{mod},\nu}^\dagger [\Psi_\epsilon]$.

The convergence between the two algebras $\mathfrak{A}_\nu^{\text{AQFT},\epsilon} (\mathcal{V}) $ and $\mathfrak{A}_\nu^{\text{mod},\epsilon} (\mathcal{V}) $ can be proved by noticing that the nonrelativistic conditions (\ref{nonrelativistic_condition_AQFT}) and (\ref{nonrelativistic_condition_mod}) are equivalent, in the sense that any function $\Psi_\epsilon (\vec{X})$ satisfying Eq.~(\ref{nonrelativistic_condition_AQFT}) satisfies Eq.~(\ref{nonrelativistic_condition_mod}) as well and the other way round. When both equations hold we have that
\begin{align}\label{A_AQFT_Psi_1_A_mod_Psi}
\hat{A}_{\text{AQFT},\nu}^\dagger [\Psi_\epsilon] = & \int_{\mathbb{R}^3} d^3X \Psi_\epsilon (\vec{X})  \hat{A}_{\text{AQFT},\nu}^\dagger(\vec{X}) \nonumber \\
\approx & \int_{\mathbb{R}^3} d^3X \Psi_\epsilon (\vec{X}) \left. \hat{A}_{\text{AQFT},\nu}^\dagger(\vec{X}) \right|_{\mathcal{H}_{\text{L},\text{R}}^\epsilon} \nonumber \\
\approx & \int_{\mathbb{R}^3} d^3X \Psi_\epsilon (\vec{X}) \left. \hat{A}_{\text{mod},\nu}^\dagger(\vec{X}) \right|_{\mathcal{H}_{\text{L},\text{R}}^\epsilon} \nonumber \\
\approx & \int_{\mathbb{R}^3} d^3X \Psi_\epsilon (\vec{X}) \hat{A}_{\text{mod},\nu}^\dagger(\vec{X}) \nonumber \\
= & \hat{A}_{\text{mod},\nu}^\dagger [\Psi_\epsilon].
\end{align}
Hence, any operator in  $\mathfrak{A}_\nu^{\text{AQFT},\epsilon} (\mathcal{V}) $ approximates to an element of $\mathfrak{A}_\nu^{\text{mod},\epsilon} (\mathcal{V}) $.

Due to the convergence between $\mathfrak{A}_\nu^{\text{AQFT},\epsilon} (\mathcal{V}) $ and $\mathfrak{A}_\nu^{\text{mod},\epsilon} (\mathcal{V}) $, we define a unified algebra $\mathfrak{A}_\nu^\epsilon (\mathcal{V}) $. Any element of $\mathfrak{A}_\nu^\epsilon (\mathcal{V}) $ is localized in $\nu$ and $\mathcal{V}$ with respect to both schemes and is nonrelativistic from the point of view of accelerated observers. We say that the localized state $| \Psi \rangle = \hat{O} | \Omega \rangle$ is nonrelativistic with respect to the Rindler frame and over the background state $| \Omega \rangle$ if $\hat{O} \in \mathfrak{A}_\nu^\epsilon (\mathcal{V}) $, which means that the state has been prepared over $| \Omega \rangle$ by the accelerated observer Rachel via the nonrelativistic local operator $\hat{O} \in \mathfrak{A}_\nu^\epsilon (\mathcal{V}) $.

\subsection{Born scheme}\label{Born_scheme_in_NRQFTCS}

In Ref.~\cite{localization_QFT}, we showed the emergence of the Born localization scheme as a result of the nonrelativistic condition (\ref{non_relativistic_limit}) in Minkowski spacetime. We demonstrated that any nonrelativistic state that is localized in any space region $\mathcal{V}$ over the Minkowski vacuum $| 0_\text{M} \rangle$ appears indistinguishable from $| 0_\text{M} \rangle$ outside $\mathcal{V}$. This results was applied to nonrelativistic ABM scenario, in which both inertial experimenters (Alice and Bob) have only access to the Minkowski bandlimited subspace $\mathcal{H}_\text{M}^\epsilon$. We found that the strict localization property  is always satisfied and, hence, the Reeh-Schlieder nonlocal effect \cite{haag1992local, Redhead1995-REDMAA-2, PhysRevA.58.135, Reeh:1961ujh} is suppressed by the nonrelativistic limit. In other words, local preparations of states by Alice never influence the measurements carried out by Bob, even when selective operations occur.

In this subsection, we show that the notion of Born localization naturally emerges in the nonrelativistic limit of Rindler spacetime [Eq.~(\ref{non_relativistic_limit_curved})] as well. In particular we demonstrate that, in such a regime, the Hilbert space $\mathcal{H}_{\text{L},\text{R}}^\epsilon$ factorizes into local Fock spaces $\mathcal{H}_\nu^\epsilon (\mathcal{V}_i)$ and the Rindler vacuum $| 0_\text{L}, 0_\text{R} \rangle $ factorizes into the local vacua $| 0_\nu (\mathcal{V}_i) \rangle $. This implies that any state satisfying Eq.~(\ref{non_relativistic_limit_curved_wavefunction}) and localized in the wedge $\nu$ and in the space region $\mathcal{V}$ over the Rindler vacuum $| 0_\text{L}, 0_\text{R} \rangle$ appears indistinguishable from $| 0_\text{L}, 0_\text{R} \rangle$ outside $\nu$ and $\mathcal{V}$. Also, the orthogonality condition for states that are localized in disjoint regions holds. We detail these results by considering the nonrelativistic RaRbR scenario and reach the same conclusions as for the ABM scenario.

The factorization of $\mathcal{H}_{\text{L},\text{R}}^\epsilon$ into local Fock spaces $\mathcal{H}_\nu^\epsilon (\mathcal{V}_i)$ can be obtained by using the definition of $\mathfrak{A}_\nu^\epsilon (\mathcal{V}) $ as the algebra generated by the operators $\hat{A}_{\text{mod},\nu} [\Psi_\epsilon]$, with $\Psi_\epsilon(\vec{X})$ supported in $\mathcal{V}$ and satisfying the nonrelativistic condition
\begin{equation}\label{A_mod_Psi_1_nonrelativistic}
\tilde{\Psi}_\epsilon (\vec{\theta}) \approx 0 \text{ if } \left| \frac{\hbar \theta_1}{mc^2} - 1 \right| \gg \epsilon,
\end{equation}
with
\begin{equation}\label{wavefunction_F_inverse_epsilon}
\tilde{\Psi}_\epsilon (\vec{\theta})  =  \frac{\sqrt{2 m c^2}}{\hbar}  \int_{\mathbb{R}^3} d^3 X  \Psi_\epsilon (\vec{X})  \mathcal{F}_\nu^*(\vec{\theta},0,\vec{X}).
\end{equation}

The commutation relation between two nonrelativistic local operators is
\begin{subequations}
\begin{align}
& \left[ \hat{A}_{\text{mod},\nu} [\Psi_\epsilon], \hat{A}_{\text{mod},\nu'}^\dagger [\Psi'_\epsilon] \right] = \delta_{\nu\nu'} \int_{\theta_1>0} d^3 \theta \tilde{\Psi}_\epsilon^* (\vec{\theta})  \tilde{\Psi}'_\epsilon (\vec{\theta}) \label{A_mod_Psi_1_commutation_a} \\
& \left[ \hat{A}_{\text{mod},\nu} [\Psi_\epsilon], \hat{A}_{\text{mod},\nu'} [\Psi'_\epsilon] \right] = 0.
\end{align}
\end{subequations}
By means of Eqs.~(\ref{FF_nu_F_nu}), (\ref{A_mod_Psi_1_nonrelativistic}) and (\ref{wavefunction_F_inverse_epsilon}), one can approximate Eq.~(\ref{A_mod_Psi_1_commutation_a}) with
\begin{align}\label{A_mod_Psi_1_commutation_a_approx}
& \left[ \hat{A}_{\text{mod},\nu} [\Psi_\epsilon], \hat{A}_{\text{mod},\nu'}^\dagger [\Psi'_\epsilon] \right] \nonumber \\
 \approx &  \delta_{\nu\nu'} \int_{| \hbar \theta_1/mc^2 - 1 | <\epsilon} d^3 \theta \tilde{\Psi}_\epsilon^* (\vec{\theta})  \tilde{\Psi}'_\epsilon (\vec{\theta}) \nonumber \\
 = & \delta_{\nu\nu'} \int_{\mathbb{R}^3} d^3 X \int_{\mathbb{R}^3} d^3 X' \int_{| \hbar \theta_1/mc^2 - 1 | <\epsilon} d^3 \theta \frac{2 m c^2}{\hbar^2} \nonumber \\
& \times  \Psi_\epsilon^* (\vec{X}) \Psi'_\epsilon (\vec{X}') \mathcal{F}_\nu(\vec{\theta},0,\vec{X}) \mathcal{F}_\nu^*(\vec{\theta},0,\vec{X}') \nonumber \\
 \approx & \delta_{\nu\nu'} \int_{\mathbb{R}^3} d^3 X \int_{\mathbb{R}^3} d^3 X' \int_{| \hbar \theta_1/mc^2 - 1 | <\epsilon} d^3 \theta \frac{2 m c^2}{\hbar^2} \nonumber \\
& \times  \Psi_\epsilon^* (\vec{X}) \Psi'_\epsilon (\vec{X}') F_\nu(\vec{\theta},0,\vec{X}) \mathcal{F}_\nu^*(\vec{\theta},0,\vec{X}') \nonumber \\
 \approx & \delta_{\nu\nu'} \int_{\mathbb{R}^3} d^3 X \int_{\mathbb{R}^3} d^3 X' \int_{\theta_1>0} d^3 \theta \frac{2 m c^2}{\hbar^2} \nonumber \\
& \times  \Psi_\epsilon^* (\vec{X}) \Psi'_\epsilon (\vec{X}') F_\nu(\vec{\theta},0,\vec{X}) \mathcal{F}_\nu^*(\vec{\theta},0,\vec{X}'),
\end{align}
which leads to
\begin{equation}\label{A_mod_Psi_1_commutation_a_approx_3}
 \left[ \hat{A}_{\text{mod},\nu} [\Psi_\epsilon], \hat{A}_{\text{mod},\nu'}^\dagger [\Psi'_\epsilon] \right] \approx \delta_{\nu\nu'} \int_{\mathbb{R}^3} d^3 X \Psi_\epsilon^* (\vec{X}) \Psi'_\epsilon ( \vec{X}),
\end{equation}
owing to
\begin{equation}\label{FF_nu_F_nu_delta_inverse}
\int_{\theta_1>0} d^3\theta  \frac{2 m c^2}{ \hbar^2 } F_\nu(\vec{\theta},0,\vec{X})  \mathcal{F}_\nu^*(\vec{\theta},0,\vec{X}') = \delta^3(\vec{X}-\vec{X}').
\end{equation}
Equation (\ref{FF_nu_F_nu_delta_inverse}) can be proven from Eqs.~(\ref{F_Rindler}), (\ref{FF_nu_F_nu}) and
\begin{equation}\label{alpha_tilde_approx_3}
\int_0^\infty d\theta_1  \frac{2 \theta_1}{\hbar a z} \tilde{F}(\vec{\theta},Z_\text{R}(z))  \tilde{F}(\vec{\theta},Z) = \frac{1}{4 \pi^2} \delta(z - z_\text{R}(Z)).
\end{equation}
A proof for Eq.~(\ref{alpha_tilde_approx_3}) is provided in Appendix \ref{appendix_2}.

Due to the localization of the wave functions $\Psi_\epsilon (\vec{X})$ and $\Psi_\epsilon' (\vec{X})$ in, respectively, $\mathcal{V}$ and $\mathcal{V}'$, we find that the right hand side of Eq.~(\ref{A_mod_Psi_1_commutation_a_approx_3}) is vanishing if $\mathcal{V}$ and $\mathcal{V}'$ are disjoint. Hence, nonrelativistic operators localized in disjoint regions commute. Consequently, the Hilbert space $\mathcal{H}_{\text{L},\text{R}}^\epsilon$ factorizes into local Hilbert spaces $\mathcal{H}_\nu^\epsilon (\mathcal{V}_i)$ upon which elements of $\mathfrak{A}_\nu^\epsilon (\mathcal{V}_i) $ act. Also, if $\Psi_\epsilon ( \vec{X})$ and $\Psi_\epsilon' ( \vec{X})$ are orthonormal with respect to the $L^2(\mathbb{R}^3)$ scalar product, then $\hat{A}_{\text{mod},\nu} [\Psi_\epsilon]$ and $\hat{A}_{\text{mod},\nu'}^\dagger [\Psi'_\epsilon]$ approximately satisfy the canonical commutation relations. This means that the local Hilbert spaces $\mathcal{H}_\nu^\epsilon (\mathcal{V})$ and $\mathcal{H}_\nu^\epsilon (\mathcal{V}')$ are also Fock spaces.

We note by $| 0_\nu (\mathcal{V}) \rangle $ the local vacuum of $\mathcal{H}_\nu^\epsilon (\mathcal{V})$. Notice that the global vacuum $| 0_\text{L}, 0_\text{R} \rangle $ is annihilated by any $ \hat{A}_{\text{mod},\nu} [\Psi_\epsilon] \in \mathfrak{A}_\nu^\epsilon (\mathcal{V})$. Hence, $| 0_\text{L}, 0_\text{R} \rangle $ factorizes into $\bigotimes_\nu \bigotimes_i | 0_\nu (\mathcal{V}_i) \rangle $ in $\mathcal{H}_{\text{L},\text{R}}^\epsilon = \bigotimes_\nu \bigotimes_i \mathcal{H}_\nu^\epsilon (\mathcal{V}) $. This means that the Rindler vacuum $| 0_\text{L}, 0_\text{R} \rangle $ is equal to $ [ \bigotimes_\nu \bigotimes_i | 0_\nu (\mathcal{V}_i) \rangle ] \otimes  \dots $ as an element of $\mathcal{H}_{\text{L},\text{R}} = \mathcal{H}_{\text{L},\text{R}}^\epsilon \otimes \dots$. Here, the dots are associated to the relativistic subspace of $\mathcal{H}_{\text{L},\text{R}}$ containing states that do not satisfy the nonrelativistic condition (\ref{non_relativistic_limit_curved}).

These results may be applied the RaRbR scenario in which a state $| \Psi \rangle = \hat{O}_\text{A} | 0_\text{L}, 0_\text{R} \rangle$ is locally prepared by an accelerated experimenter (Rachel) over the Rindler vacuum background $| 0_\text{L}, 0_\text{R} \rangle$ and a second accelerated observer (Rob) performs local measurements of the observable $\hat{O}_\text{B}$. The operators $ \hat{O}_\text{A}$ and $ \hat{O}_\text{B}$ are, respectively, localized in the wedges $\nu_\text{A}$ and $\nu_\text{B}$ and in the space regions $\mathcal{V}_\text{A}$ and $\mathcal{V}_\text{B}$. We assume that the state prepared by Rachel and the observable measured by Rob are nonrelativistic with respect to the Rindler frame, which means that $\hat{O}_\text{A} \in \mathfrak{A}_{\nu_\text{A}}^\epsilon (\mathcal{V}_\text{A}) $ and $\hat{O}_\text{B} \in \mathfrak{A}_{\nu_\text{B}}^\epsilon (\mathcal{V}_\text{B}) $.

As a consequence of the factorizations $\mathcal{H}_{\text{L},\text{R}}^\epsilon = \bigotimes_\nu \bigotimes_i \mathcal{H}_\nu^\epsilon (\mathcal{V}_i) $ and $| 0_\text{L}, 0_\text{R} \rangle  = [ \bigotimes_\nu \bigotimes_i | 0_\nu (\mathcal{V}_i) \rangle ] \otimes  \dots  $, the strict localization property is always satisfied in the nonrelativistic RaRbR scenario. In particular, we find that
\begin{equation}
\langle \Psi | \hat{O}_\text{B} | \Psi \rangle = \langle 0_\text{L}, 0_\text{R} | \hat{O}_\text{B}  | 0_\text{L}, 0_\text{R} \rangle
\end{equation}
when $\nu_\text{A}$ is different from $\nu_\text{B}$ or when $\mathcal{V}_\text{A}$ and $\mathcal{V}_\text{B}$ are disjoint. This means that Rob's measurements are never influenced by Rachel's preparations of states.

Such a result is in analogy to what we found for the ABM scenario. We conclude that nonlocal effects are always suppressed if the nonrelativistic limit and the background vacuum state are associated to the same frame.

\subsection{Nonrelativistic RaRbM scenario} \label{The_modal_scheme_over_general_background_does_not_converge_to_the_Born_scheme}

In this subsection, we study the RaRbM scenario in the nonrelativistic limit. At variance with the nonrelativistic RaRbR scenario considered in Sec.~\ref{Born_scheme_in_NRQFTCS}, here, the background state is the Minkowski vacuum $| 0_\text{M} \rangle$. Hence, the state prepared by Rachel is $| \Psi \rangle = \hat{O}_\text{A} | 0_\text{M} \rangle$ with $\hat{O}_\text{A} \in \mathfrak{A}_{\nu_\text{A}}^\epsilon (\mathcal{V}_\text{A}) $. The observable measured by Rob, instead, is still $\hat{O}_\text{B} \in \mathfrak{A}_{\nu_\text{B}}^\epsilon (\mathcal{V}_\text{B}) $.

In Ref.~\cite{localization_QFT}, we showed that the nonrelativistic Minkowski-Fock space $\mathcal{H}_\text{M}^\epsilon$ factorizes into local Fock spaces $\mathcal{H}_\text{M}^\epsilon(\mathcal{V}_i)$ with vacuum states $| 0_\text{M} (\mathcal{V}_i) \rangle$ and the Minkowski vacuum $| 0_\text{M} \rangle$ factorizes into $\bigotimes_i | 0_\text{M} (\mathcal{V}_i) \rangle$ with respect to $\mathcal{H}_\text{M}^\epsilon = \bigotimes_i \mathcal{H}_\text{M}^\epsilon(\mathcal{V}_i)$. However, here, we are interested in the factorization with respect to $\mathcal{H}_{\text{L},\text{R}}^\epsilon = \bigotimes_\nu \bigotimes_i \mathcal{H}_\nu^\epsilon (\mathcal{V}_i) $.

The representation of the Minkowski vacuum $| 0_\text{M} \rangle$ in the Rindler-Fock space $\mathcal{H}_{\text{L},\text{R}}$ is given by \cite{RevModPhys.80.787}
\begin{align} \label{SS_S_31}
| 0_\text{M} \rangle \propto & \exp \left( \sum_{\nu=\{\text{L},\text{R}\}} \int_0^\infty d\Omega \int_{\mathbb{R}^2} d^2\vec{K}_\perp e^{-\pi \Omega/ca} \right. \nonumber \\
& \left. \times \hat{A}^\dagger_\nu(\Omega,\vec{K}_\perp) \hat{A}^\dagger_{\bar{\nu}}(\Omega,-\vec{K}_\perp) \right) | 0_\text{L}, 0_\text{R} \rangle,
\end{align}
with $\bar{\nu}$ as the opposite of $\nu$, i.e., $\bar{\nu}=\text{L} $ if $\nu= \text{R}$ and $\bar{\nu}= \text{R}$ if $\nu= \text{L}$. As a consequence of Eq.~(\ref{SS_S_31}) and due to the factorization of the Rindler vacuum $| 0_\text{L}, 0_\text{R} \rangle  = [\bigotimes_\nu \bigotimes_i | 0_\nu (\mathcal{V}_i) \rangle] \otimes \dots $, we find that the Minkowski vacuum $| 0_\text{M} \rangle$ is entangled between the nonrelativistic local Fock spaces $\mathcal{H}_\nu^\epsilon (\mathcal{V}_i) $.

As a consequence of this entanglement, we find that, in general,
\begin{equation}
\langle \Psi | \hat{O}_\text{B} | \Psi \rangle \neq \langle 0_\text{M} | \hat{O}_\text{B}  | 0_\text{M} \rangle
\end{equation}
even if $\mathcal{V}_\text{A}$ and $\mathcal{V}_\text{B}$ are disjoint. However, the identity
\begin{equation}
\langle \Psi | \hat{O}_\text{B} | \Psi \rangle = \langle 0_\text{M} | \hat{O}_\text{B}  | 0_\text{M} \rangle
\end{equation}
is guaranteed when $\hat{O}_\text{A}$ is unitary and either the two wedges $\nu_\text{A} $ and $ \nu_\text{B}$ are different or the two regions $\mathcal{V}_\text{A}$ and $\mathcal{V}_\text{B}$ are disjoint. This is due the fact that $\hat{O}_\text{A}$ and $\hat{O}_\text{B}$ commute and, hence, Eq.~(\ref{local_unitary_operator_measurament_A}) holds.

As a result, we find that the strict localization property in the nonrelativistic RaRbM scenario is guaranteed for unitary preparations of the state. By only performing nonselective operations on the vacuum $|0_\text{M} \rangle$, Rachel does not influence Rob's measurements in the other disjoint region. Conversely, selective preparations of the state lead to nonlocal effects.

The nonlocality shown here is similar to the Reeh-Schlieder effect \cite{haag1992local, Redhead1995-REDMAA-2, PhysRevA.58.135, Reeh:1961ujh}. Indeed, the origin of this effect is ascribed to the entanglement between local spaces induced by the background vacuum state. These correlations are suppressed in the nonrelativistic limit with respect to the Minkowski frame. However, here, we are considering the nonrelativistic regime of accelerated observers and we find that no suppression occurs in the RaRbM scenario.

\subsection{Nonrelativistic ARbM scenario} \label{AliceRob_nonlocality_is_not_suppressed_by_the_nonrelativistic_limit}

In Sec.~\ref{Born_scheme_in_NRQFTCS}, we considered the nonrelativistic ABM scenario, in which all experimenters are inertial and the background state is the Minkowski vacuum $| 0_\text{M} \rangle$. In that case, physical phenomena are described by means of the Hilbert space $\mathcal{H}_\text{M}^\epsilon$ and the nonrelativistic local algebra  $\mathfrak{A}_\text{M}^\epsilon(\mathcal{V})$, which is defined as the subalgebra of $\mathfrak{A}_\text{M}^\text{mod}(\mathcal{V})$ generated by operators of the form of
\begin{equation}
\hat{a}_\text{mod}^\dagger [\psi_\epsilon] = \int_{\mathbb{R}^3} d^3x \psi_\epsilon (\vec{x})  \hat{a}_\text{mod}^\dagger(\vec{x}),
\end{equation}
where $\psi_\epsilon (\vec{x})$ is supported in $\mathcal{V}$ and its Fourier transform $\tilde{\psi}_\epsilon (\vec{k})$ is supported in the nonrelativistic region (\ref{non_relativistic_limit}). Due to the convergence between the different localization schemes in the nonrelativistic limit, one can use $\mathfrak{A}_\text{M}^\text{AQFT}(\mathcal{V})$ and $\hat{a}_\text{AQFT}^\dagger(\vec{x})$ instead of $\mathfrak{A}_\text{M}^\text{mod}(\mathcal{V})$ and $ \hat{a}_\text{mod}^\dagger(\vec{x})$ to define $\mathfrak{A}_\text{M}^\epsilon(\mathcal{V})$. Any state $| \psi \rangle$ of the nonrelativistic Hilbert space $\mathcal{H}_\text{M}^\epsilon$, localized in $\mathcal{V}_\text{A}$ can be identified by an operator $\hat{O}_\text{A} \in \mathfrak{A}_\text{M}^\epsilon(\mathcal{V}_\text{A})$ such that $| \psi \rangle = \hat{O}_\text{A} | 0_\text{M} \rangle$.

At variance with Sec.~\ref{Born_scheme_in_NRQFTCS}, here, we consider the nonrelativistic ARbM scenario, in which the observer performing the measurement (i.e., experimenter B) is accelerated (Rob). Hence, the operator $\hat{O}_\text{B}$ is an element of $\mathfrak{A}_{\nu_\text{B}}^\epsilon(\mathcal{V}_\text{B})$ instead of $\mathfrak{A}_\text{M}^\epsilon(\mathcal{V}_\text{B})$, which means that $\hat{O}_\text{B}$ is a linear combination of products of $\hat{A}_{\text{mod},\nu_\text{B}} [\Psi_\epsilon]$ operators, with $\Psi_\epsilon(\vec{X})$ supported in $\mathcal{V}_\text{B}$ and satisfying the nonrelativistic condition in the Rindler frame (\ref{A_mod_Psi_1_nonrelativistic}).

In Ref.~\cite{PhysRevD.107.085016}, we showed that the nonrelativistic condition in the two frames are not compatible. Intuitively, this can be seen by noticing that inertial and accelerated observers experience different flows of times due to the inequivalent time coordinates $t$ and $T$; hence, they are provided with different notions of energy as the generator of time translation. This means that the nonrelativistic condition appears to be frame dependent. Consequently, if $\Psi(\vec{X})$ satisfies the nonrelativistic condition in the Rindler frame (\ref{A_mod_Psi_1_nonrelativistic}), then, for any $\mathcal{V}'_\text{B} \subseteq \mathbb{R}^3$, the operator $\hat{A}_{\text{mod},\nu} [\Psi_\epsilon]$ is not an element of the Minkowski nonrelativistic local algebra $\mathfrak{A}_\text{M}^\epsilon(\mathcal{V}'_\text{B})$. In other words, the Minkowski and the Rindler nonrelativistic schemes are incompatible.

Due to the incompatibility between the two schemes, the strict localization property
\begin{equation}\label{KnightLicht_property}
\langle \phi | \hat{O}_\text{B} | \phi \rangle = \langle 0_\text{M} | \hat{O}_\text{B}  | 0_\text{M} \rangle.
\end{equation}
does not generally hold. This means that the preparation of the state $| \psi \rangle = \hat{O}_\text{A} |0_\text{M} \rangle $ by Alice, with $\hat{O}_\text{A} \in \mathfrak{A}_\text{M}^\epsilon(\mathcal{V}_\text{A})$, may influence the measurement of $\hat{O}_\text{B}$ by Rob, even if $\mathcal{V}_\text{A}$ and $\mathcal{V}_\text{B}$ represent different regions of the spacetime.

However, unitary preparations of $| \psi \rangle$ guarantee the independence between the local preparation of the state and the local measurements of the observable $\hat{O}_\text{B}$. This is a consequence of the microcausality axiom that ensures that operators localized in different region of spacetimes with respect to the AQFT scheme commute. Since the algebras $\mathfrak{A}_\text{M}^\epsilon(\mathcal{V}_\text{A})$ and $\mathfrak{A}_{\nu_\text{B}}^\epsilon(\mathcal{V}_\text{B})$ are subalgebras of, respectively, $\mathfrak{A}_\text{M}^\text{AQFT}(\mathcal{V}_\text{A})$ and $\mathfrak{A}_{\nu_\text{B}}^\text{AQFT}(\mathcal{V}_\text{B})$, we find that $\hat{O}_\text{A}$ and $\hat{O}_\text{B}$ commute if $\mathcal{V}_\text{A}$ and $\mathcal{V}_\text{B}$ represent different regions of the spacetime. Then, one can use Eq.~(\ref{local_unitary_operator_measurament_A}) to prove that, if $\hat{O}_\text{A}$ is unitary, Eq.~(\ref{KnightLicht_property}) holds.

In conclusion, we find that, at variance with the ABM and the RaRbR scenarios, in the ARbM scenario the Reeh-Schlieder nonlocal effects are not suppressed. This is a consequence of the incompatibility between the nonrelativistic limit in the two frames and the consequent incompatibility between the respective nonrelativistic localization schemes.

\section{Conclusions}\label{Localization_in_accelerated_frame_Conclusions}

The algebraic approach to QFTCS provides an exact description of local physical phenomena in a regime in which relativistic energies and noninertial effects cannot be ignored. At variance with the modal representation of particle states, the AQFT localization scheme is frame independent and does not violate relativistic causality. Only in the nonrelativistic regime the two schemes appear indistinguishable from each other.

Notwithstanding the relativistic causal nature of the AQFT scheme, nonlocal instantaneous effects occur when states are selectively prepared over any background $| \Omega \rangle$ in confined regions $\mathcal{V}_\text{A}$. In particular, local observations in regions $\mathcal{V}_\text{B}$ disjoint from $\mathcal{V}_\text{A}$ are influenced by the nonunitary preparations of states in $\mathcal{V}_\text{A}$. It is known that such a nonlocal effect is suppressed in the nonrelativistic limit of Minkowski spacetime, with $| \Omega \rangle$ being equal to the Minkowski vacuum $| 0_\text{M} \rangle$ \cite{localization_QFT}. Here, we reach the conclusion that the suppression generally occurs in the nonrelativistic regime of the frame associated to the vacuum background $| \Omega \rangle$.

Different frames are characterized by different time coordinates, which lead to different notions of vacuum state and nonrelativistic limit \cite{PhysRevD.107.085016}. Hence, it is not surprising that for any (inertial or accelerated) observer there is a preferred background state ($| 0_\text{M} \rangle$ or $| 0_\text{L}, 0_\text{R} \rangle$) and a preferred nonrelativistic condition (Eq.~(\ref{non_relativistic_limit}) or Eq.~(\ref{non_relativistic_limit_curved})). Nonlocal effects are completely suppressed when all of these frame dependent elements match up, i.e., when the vacuum background state and the nonrelativistic condition are associated to the same frame. The nonrelativistic ABM and the nonrelativistic RaRbR scenarios fall into this category and are associated to, respectively, the Minkowski and the Rindler frame.

In the RaRbM and the ARbM scenarios, instead, the nonlocal effects are not suppressed, due to the aforementioned elements not matching up. Specifically, in the RaRbM scenario, the background state is the Minkowski vacuum $| 0_\text{M} \rangle$, whereas the nonrelativistic condition is associated to the Rindler frame. The origin of the nonlocal effect needs to be ascribed to the entanglement of $| 0_\text{M} \rangle$ between the Rindler nonrelativistic local Fock spaces. Conversely, in the ARbM scenario, states are prepared by the inertial observer Alice via nonrelativistic Minkowski operators over the Minkowski vacuum $| 0_\text{M} \rangle$; whereas, local measurements are performed by the accelerated observer Rob via nonrelativistic Rindler operators. From Alice's point of view, Rob uses relativistic observables, since her notion of nonrelativistic limit is different than Rob's. Hence, the Reeh-Schlieder nonlocality is not suppressed.

These theoretical results may find practical applications in the context of, e.g., nonrelativistic emitters and detectors. If both instruments are inertial and operate over the Minkowski vacuum $| 0_\text{M} \rangle$, no nonlocal effect is expected to be detected. The same occurs if they are both accelerated over the Rindler vacuum background $| 0_\text{L}, 0_\text{R} \rangle$. Conversely, if at least one of them is accelerated and the background state is the Minkowski vacuum $| 0_\text{M} \rangle$, then nonlocal effects may appear and can be eventually measured.

By means of this nonlocality, it is possible to probe the nature of the background state in the presence of gravitational fields (e.g., Earth gravity). Due to the Einstein's equivalence principle, any standing experimenter affected by a gravitational field can be locally represented by the Rindler frame, whereas a free falling observer is described by the Minkowski frame. In this scenario, we still do not have experimental evidence about the nature of the background state: is it a Minkowski or a Rindler vacuum? To answer this question, one may use nonrelativistic emitters and detectors that follow different trajectories and measure eventual nonlocal effects.

\section*{Acknowledgment}

We acknowledge financial support from CN1 Quantum PNRR MUR CN 0000013 HPC and by the HORIZON-EIC-2022-PATHFINDERCHALLENGES-01 HEISINGBERG project 101114978.

\appendix

\section{} \label{Proof_of_alpha_tilde_approx_3_inverse}

In this section we provide a proof for Eq.~(\ref{alpha_tilde_approx_3_inverse}). The method is based on the use of the Kontorovich–Lebedev, defined as
\begin{equation}\label{KontorovichLebedev}
\mathcal{K} [\varphi] (\zeta) = \frac{2 \zeta}{\pi^2} \sinh(\pi \zeta) \int_0^\infty \frac{d\xi}{\xi}  K_{i \zeta}(\xi)  \varphi(\xi)
\end{equation}
for any function $\varphi(\xi)$ in $\xi>0$. The inverse of Eq.~(\ref{KontorovichLebedev}) is
\begin{equation}
\mathcal{K}^{-1} [\varphi] (\xi) = \int_0^\infty d\zeta  K_{i \zeta}(\xi)  \varphi(\zeta).
\end{equation}

We consider any function $\varphi(\zeta)$ with $\zeta>0$ and we compute the following integral
\begin{align} \label{Kontorovich_Lebedev_2_0}
&\int_0^\infty d\Omega \int_0^\infty \frac{dx  2 \Omega'}{\hbar a z} \sqrt{\frac{\sinh \left( \frac{\beta \Omega'}{2} \right)}{\sinh \left( \frac{\beta \Omega}{2} \right)}}  \nonumber \\
&  \times  \tilde{F}(\Omega, \vec{K}_\perp,Z_\text{R}(z)) \tilde{F}(\Omega', \vec{K}_\perp,Z_\text{R}(z)) \varphi\left( \frac{\Omega}{ca} \right)\nonumber \\
 = & \frac{1}{2\pi^4 (c a)^2} \int_0^\infty d\Omega \int_0^\infty dx  \frac{\Omega'}{z} \sinh \left( \frac{\beta \Omega'}{2} \right)   \nonumber \\
&  \times  K_{i \Omega /ca} \left( \kappa(\vec{K}_\perp) z \right) K_{i \Omega' /ca} \left( \kappa(\vec{K}_\perp) z \right)   \varphi\left(  \frac{\Omega}{ca}\right).
\end{align}
We consider the coordinate transformation
\begin{align}
& \zeta = \frac{\Omega}{ca}, & \xi = \kappa(\vec{K}_\perp) z,
\end{align}
which leads to
\begin{align} \label{Kontorovich_Lebedev_2_1}
& \int_0^\infty d\Omega \int_0^\infty \frac{dx  2 \Omega'}{\hbar a z} \sqrt{\frac{\sinh \left( \frac{\beta \Omega'}{2} \right)}{\sinh \left( \frac{\beta \Omega}{2} \right)}}  \nonumber \\
&  \times  \tilde{F}(\Omega, \vec{K}_\perp,Z_\text{R}(z)) \tilde{F}(\Omega', \vec{K}_\perp,Z_\text{R}(z)) \varphi\left( \frac{\Omega}{ca} \right)\nonumber  \\
=  &  \frac{1}{2 \pi^4 ca}  \int_0^\infty d\xi \int_0^\infty d\zeta \frac{\Omega'}{ \xi} \sinh \left( \frac{\beta \Omega'}{2}  \right)  \nonumber \\
&  \times  K_{i \zeta} (\xi)  K_{i \Omega'/ca} (\xi) \varphi (\zeta)\nonumber \\
= & \frac{1}{4 \pi^2}  \mathcal{K} [\mathcal{K}^{-1} [\varphi]] \left( \frac{\Omega'}{ca} \right)\nonumber \\
= & \frac{1}{4 \pi^2}  \varphi \left( \frac{\Omega'}{ca} \right).
\end{align}
By using the generality of $\varphi(\zeta)$, we prove Eq.~(\ref{alpha_tilde_approx_3_inverse}).

\section{} \label{appendix_2}

By using a method similar to the one adopted in Appendix \ref{Proof_of_alpha_tilde_approx_3_inverse}, here, we demonstrate Eq.~(\ref{alpha_tilde_approx_3}). In this case, we consider a function $\varphi(\xi)$ in $\xi>0$ and we compute the integral
\begin{align} \label{Kontorovich_Lebedev_0}
&\int_0^\infty dx \int_0^\infty d\theta_1  \frac{2 \theta_1}{\hbar a z}  \tilde{F}(\vec{\theta},Z_\text{R}(z)) \tilde{F}(\vec{\theta},Z) \varphi\left( \kappa(\vec{\theta}_\perp) z \right)\nonumber \\
 = & \frac{1}{2\pi^4 (c a)^2} \int_0^\infty dx \int_0^\infty d\theta_1  \frac{\theta_1}{z} \sinh \left( \frac{\beta \theta_1}{2} \right)   \nonumber \\
&  \times K_{i \theta_1 /ca} \left( \kappa(\vec{\theta}_\perp) z \right)  K_{i \theta_1 /ca} \left( \kappa(\vec{\theta}_\perp) \frac{e^{aX}}{a} \right)   \varphi\left( \kappa(\vec{\theta}_\perp) z \right).
\end{align}
Here, we use the coordinate transformation
\begin{align}
& \zeta = \frac{\theta_1}{ca}, & \xi = \kappa(\vec{\theta}_\perp) z,
\end{align}
which gives
\begin{align} \label{Kontorovich_Lebedev_1}
&\int_0^\infty dx \int_0^\infty d\theta_1  \frac{2 \theta_1}{\hbar a z}  \tilde{F}(\vec{\theta},Z_\text{R}(z)) \tilde{F}(\vec{\theta},Z)  \varphi\left( \kappa(\vec{\theta}_\perp) z \right) \nonumber \\
=  &  \frac{1}{2 \pi^4} \int_0^\infty d\xi \int_0^\infty d\zeta \frac{\zeta}{ \xi} \sinh ( \pi \zeta )  K_{i \zeta} (\xi)  K_{i \zeta} \left( \kappa(\vec{\theta}_\perp) \frac{e^{aX}}{a} \right) \nonumber \\
&  \times  \varphi (\xi) \nonumber \\
= & \frac{1}{4 \pi^2}  \mathcal{K}^{-1} [\mathcal{K} [\varphi]] \left( \kappa(\vec{\theta}_\perp) \frac{e^{aX}}{a} \right)\nonumber \\
 = & \frac{1}{4 \pi^2}  \varphi \left( \kappa(\vec{\theta}_\perp) \frac{e^{aX}}{a} \right).
\end{align}
Equation (\ref{Kontorovich_Lebedev_1}) proves Eq.~(\ref{alpha_tilde_approx_3}) as it holds for any $\varphi(\xi)$.

\bibliography{bibliography}

\end{document}